\documentclass[11pt]{article}
\usepackage{amssymb,amsmath,cite}
\catcode`\@=11

\@addtoreset{equation}{section}
\@addtoreset{equation}{subsection}
\def\theequation{\arabic{section}.\arabic{subsection}.\arabic{equation}}
     
\catcode`\@=11
\def\thesection{\arabic{section}}

\def\appendix{\setcounter{section}{0}
        \def\thesection{Appendix \Alph{section}}
        \def\theequation{\Alph{section}.\arabic{equation}}}
\def\section{\@startsection{section}{1}{\z@}{3.5ex plus 1ex minus
   .2ex}{2.3ex plus .2ex}{\large\bf}}
     
\long\def\@makefntext#1{\parindent 0cm\noindent
\hbox to 1em{\hss$^{\@thefnmark}$}#1}

\def\IR{{\hbox{{\rm I}\kern-.2em\hbox{\rm R}}}}
\def\IH{{\hbox{{\rm I}\kern-.2em\hbox{\rm H}}}}
\def\IC{{\ \hbox{{\rm I}\kern-.6em\hbox{\bf C}}}}
\def\IZ{{\hbox{{\rm Z}\kern-.4em\hbox{\rm Z}}}}
\def\rref#1{(\ref{#1})}
\newcommand{\beq}{\begin{equation}}
\newcommand{\eeq}{\end{equation}}

\begin{document}
\begin{titlepage}
\vspace{.5in}
\begin{flushright}
UCD-05-02\\
gr-qc/0503022\\
March 2005\\
\end{flushright}
\vspace{.5in}

\begin{center}
{\Large\bf  
Conformal Field Theory,\\[1.8ex]    
(2+1)-Dimensional Gravity, and the BTZ Black Hole}\\

\vspace{.4in}
{S.~C{\sc arlip}\footnote{\it email: carlip@physics.ucdavis.edu}\\
       {\small\it Department of Physics}\\
       {\small\it University of California}\\
       {\small\it Davis, CA 95616}\\{\small\it USA}}
\end{center}

\vspace{.5in}
\begin{center}
{\large\bf Abstract}
\end{center}
\begin{center}
\begin{minipage}{4.75in}
{\small In three spacetime dimensions, general relativity becomes 
a topological field theory, whose dynamics can be largely described 
holographically by a two-dimensional conformal field theory at the 
``boundary'' of spacetime.  I review what is known about this reduction%
---mainly within the context of pure (2+1)-dimensional gravity---and 
discuss its implications for our understanding of the statistical 
mechanics and quantum mechanics of black holes.
}
\end{minipage}
\end{center}
\end{titlepage}
\addtocounter{footnote}{-1}

It has been thirty years since Hawking first showed that black holes
were thermodynamic objects, with characteristic temperatures and
entropies \cite{Hawking}.  For most of that time---and, indeed, even
before Hawking's work \cite{Bekenstein}---it has been assumed that
these thermodynamic properties reflect the statistical mechanics of
underlying quantum gravitational states.  But the detailed nature 
of these states has remained a mystery.  The recent proliferation of
state-counting methods, in string theory \cite{Strominger}, AdS/CFT
\cite{AGMOO}, loop quantum gravity \cite{Baez}, and induced
gravity \cite{Frolov}, has, if anything, deepened the mystery: we must
now also explain the ``universality'' of black hole entropy \cite{Carlipa},
the fact that so many distinct and seemingly orthogonal approaches reach 
the same conclusion.

The problem of black hole statistical mechanics is especially stark
in (2+1)-dimensional spacetime.  In three spacetime dimensions, general 
relativity becomes a topological field theory \cite{Martinec,Achucarro,%
Witten,Wittena} with only a few, nonpropagating degrees of freedom;
there seems to be little room for quantum states that might 
account for black hole thermodynamics.  The (2+1)-dimensional BTZ black 
hole of Ba{\~n}ados, Teitelboim, and Zanelli \cite{BTZ,BHTZ}, however, 
can have an arbitrarily high entropy.  If we can explain this entropy in 
such a simple setting, it may take us a long way towards understanding 
the general problem.  This lower-dimensional model becomes even more 
interesting when one notes that the near-extremal black holes whose 
entropy can be computed in string theory almost all have a near-horizon 
geometry containing the BTZ solution, and that the corresponding 
entropies can be determined from this dimensionally reduced geometry.

In this review, I will summarize the current---highly incomplete---%
understanding of the microscopic statistical mechanics of the BTZ black 
hole.  The key will be that much of (2+1)-dimensional gravity can be 
described ``holographically'' by a two-dimensional conformal field theory.  
This conformal field theory is, unfortunately, of a type that is still 
poorly understood, so many questions remain.  But considerable progress 
has now been made, and some directions for further research are clear.

The literature in this field is by now enormous, and my treatment 
will necessarily be rather incomplete.  This paper should be read as a  
personal overview, not as a comprehensive review.  In particular, I will 
have relatively little to say about string theory, and will only touch
briefly on the extension to supergravity.  Several sections of this 
work are based on a previous paper, Ref.\ \cite{Carlipg}.

\section{(2+1)-Dimensional Gravity \label{tog}}

I will begin with a very brief summary of some key aspects of general 
relativity in three spacetime dimensions.  Much more extensive discussions 
can be found in \cite{Carlipb,Carlipc,Carlipd}.  The idea that (2+1)-%
dimensional gravity might be a useful arena for investigating more general
questions dates back to at least 1966 \cite{Leut}, but the power of the 
model only became clear with the work of Deser, Jackiw, and 't Hooft 
\cite{DJtH,DJ,tH,DJb}.  For us, the first essential feature is that (2+1)-%
dimensional vacuum gravity has no local degrees of freedom \cite{Star,Leut}.  
This can be shown by a simple counting argument: the phase space consists 
of a spatial metric (three degrees of freedom per point) and its canonical 
momentum (another three degrees of freedom per point); but the theory also 
has three constraints that restrict initial data and three arbitrary coordinate 
choices, leaving no unconstrained, non-``gauge'' degrees of freedom.

Alternatively, one can note that the curvature tensor in 2+1 dimensions 
is algebraically determined by the Ricci tensor:
\beq
G^\mu{}_\nu = -\frac{1}{4}\epsilon^{\mu\pi\rho}\epsilon_{\nu\sigma\tau}
  R_{\pi\rho}{}^{\sigma\tau} .
\label{a0}
\eeq
In particular, a vacuum 
solution ($R_{\mu\nu}=0$) necessarily has a vanishing curvature, and can 
therefore be constructed by ``gluing together'' flat pieces of Minkowski 
space.  For a topologically nontrivial manifold, this gluing need not 
be unique, and one can have a variety of inequivalent ``geometric 
structures'' \cite{Carlipd,Carlipe}.  But these are labeled by a finite 
number of global parameters that describe the gluing, and do not involve
any propagating degrees of freedom.  Similarly, any solution of the vacuum 
field equations with a negative cosmological constant has constant negative 
curvature, and can be constructed by gluing patches of anti-de Sitter space.

For spacetimes with boundaries or asymptotic regions, this picture becomes 
a bit more complex.  For the action principle to hold---that is, for the
Einstein-Hilbert action to have any extrema at all---one must typically
introduce boundary conditions on the fields and add boundary terms to the 
action.  These generically break the gauge and diffeomorphism symmetries
of the theory: configurations that are gauge equivalent in the absence of
boundaries may not be connected by transformations that behave properly at
the boundaries.  Moreover, the remaining transformations at the boundary are 
properly viewed as symmetries, not gauge invariances \cite{RegTeit,Benguria}:
as explained below in section \ref{GIS}, physical states need not be 
invariant, but can transform under representations of the group of boundary 
transformations.  As a consequence, new ``would-be pure gauge'' degrees of 
freedom appear at the boundary \cite{Carlipa}.

At first sight, this treatment of boundary gauge transformations may
seem somewhat arbitrary.  But in certain cases, including Chern-Simons
theory---which, as we shall see below, is closely linked to (2+1)-dimensional 
gravity---the ``would-be pure gauge'' degrees of freedom are needed to
provide a complete intermediate set of quantum states \cite{Wittenb}.  In 
particular, consider a Chern-Simons path integral on a manifold $M_1$ with 
boundary $\Sigma$.  The partition function $Z_{(M_1,\Sigma)}[A,g]$ depends
on the boundary value of the gauge field $A$ and on a gauge parameter $g$
that acts as a propagating field on $\Sigma$.  If $M_2$ is another manifold
with a diffeomorphic boundary $\Sigma$, one can ``glue'' $M_1$ and $M_2$
along $\Sigma$ to obtain a compact manifold $M = M_1\cup_\Sigma M_2$.  The
path integral equivalent of summing over intermediate states is to set the 
boundary values of the fields equal and integrate over these boundary fields:
\beq
Z_M = \int[dA][dg]Z_{(M_1,\Sigma)}[A,g]Z_{(M_2,\Sigma)}[A,g] .
\label{a1}
\eeq
Witten shows in \cite{Wittenb} that this process gives the correct Chern-Simons
partition function for $M$ only if the ``gauge'' degrees of freedom $g$ are
included, with the proper Wess-Zumino-Witten action described below in section 
\ref{CSWZW}.  Thus, at least for Chern-Simons theory, the ``would-be gauge'' 
degrees of freedom are \emph{necessary} for a consistent quantum theory.

\subsection{Gravity and Chern-Simons Theory}

As first noted by Achucarro and Townsend \cite{Achucarro} and subsequently
extensively developed by Witten \cite{Witten,Wittena}, vacuum Einstein gravity 
in three spacetime dimensions is equivalent to a Chern-Simons gauge theory.
We will be interested in the case of a negative cosmological constant
$\Lambda = -1/\ell^2$.  Then the coframe (or ``triad'' or ``dreibein'') 
$e^a = e_\mu{}^adx^\mu$ and the spin connection $\omega^a = \frac{1}{2}
\epsilon^{abc} \omega_{\mu bc}dx^\mu$ can be combined into two $\mathrm{SL}
(2,\mathbb{R})$ connection one-forms
\beq
A^{(\pm)a} = \omega^a \pm \frac{1}{\ell}e^a .
\label{aa1}
\eeq
It is straightforward to show that up to possible boundary terms,
the first-order form of the usual Einstein-Hilbert action can be written 
as
\beq
I = \frac{1}{8\pi G}\int_M \left\{ e^a\wedge\left( d\omega_a 
  + \frac{1}{2}\epsilon_{abc} \omega^b\wedge \omega^c \right)
  + \frac{\Lambda}{6}\epsilon_{abc} e^a\wedge e^b\wedge e^c \right\}
  = I_{\mathit{CS}}[A^{(+)}] 
  - I_{\mathit{CS}}[A^{(-)}]
\label{aa2}
\eeq
where $A^{(\pm)} = A^{(\pm)a}T_a$ are $\mathrm{SL}(2,\mathbb{R})$-valued 
gauge potentials (see \ref{App} for conventions for the generators
$T_a$), and the Chern-Simons action $I_{\mathit{CS}}$ is
\beq
I_{\mathit{CS}} = \frac{k}{4\pi} \int_M \mathrm{Tr}
  \left\{ A\wedge dA + \frac{2}{3} A\wedge A\wedge A \right\} ,
\label{aa3}
\eeq
with 
\beq
k=\frac{\ell}{4G} .
\label{aa3a}
\eeq  
Similarly, the Chern-Simons field equations
\beq
F^{(\pm)} = dA^{(\pm)} + A^{(\pm)}\wedge A^{(\pm)} = 0
\label{aa4}
\eeq
are easily seen to be equivalent to the requirement that the
connection be torsion-free and that the metric have constant negative 
curvature, as required by the vacuum Einstein field equations.

This formulation has the enormous advantage that gravity becomes an 
ordinary gauge theory.  In particular, diffeomorphisms, which create 
endless complications in standard approaches to quantum gravity, are
now equivalent on shell to ordinary gauge transformations.  Indeed,
the Lie derivative of the connection is
\beq
\mathcal{L}_\xi A = d\left(\xi\cdot A\right) + \xi\cdot dA
  = \xi\cdot F + D_A\left(\xi\cdot A\right)
\label{aa5}
\eeq
where $D_A$ is the gauge-covariant exterior derivative.  It is
evident that on shell---that is, when $F=0$---eqn.~(\ref{aa5}) is 
simply an ordinary infinitesimal gauge transformation with gauge 
parameter $\lambda^a = \xi^\mu A_\mu{}^a$.

\subsection{The BTZ Black Hole \label{btz}}

When the cosmological constant is zero, a vacuum solution of (2+1)-%
dimensional gravity is necessarily flat, and it can be shown that there
are no black hole solutions \cite{BH}.  It therefore came as an enormous
surprise when Ba{\~n}ados, Teitelboim, and Zanelli showed that vacuum
(2+1)-dimensional gravity with $\Lambda<0$ admitted a black hole solution 
\cite{BTZ}.  A review of this solution is given in \cite{Carlipf}; here
I will just touch on a few of the most relevant features.

The BTZ black hole in ``Schwarzschild'' coordinates is given by the 
metric
\beq
ds^2 = (N^\perp)^2dt^2 - f^{-2}dr^2
  - r^2\left( d\phi + N^\phi dt\right)^2
\label{ab1}
\eeq
with lapse and shift functions and radial metric\footnote{Many papers
use the conventions of \cite{BTZ}, in which units are chosen such that 
$8G=1$.}
\beq
N^\perp = f
  = \left( -8GM + \frac{r^2}{\ell^2} + \frac{16G^2J^2}{r^2} \right)^{1/2} ,
  \quad N^\phi = - \frac{4GJ}{r^2} \qquad  (|J|\le M\ell) .
\label{ab2}
\eeq
The metric \rref{ab1} is stationary and axially symmetric, with Killing 
vectors $\partial_t$ and $\partial_\phi$, and generically has no other 
symmetries.  Although it describes a spacetime of constant negative
curvature, it is a true black hole: it has a genuine event horizon at
$r_+$ and, when $J\ne0$, an inner Cauchy horizon at $r_-$, where
\beq
r_\pm^2=4GM\ell^2\left \{ 1 \pm
\left [ 1 - \left(\frac{J}{M\ell}\right )^2\right ]^{1/2}\right \} ,
\label{ab3}
\eeq
i.e.,
\beq
M=\frac{r_+^2+r_-^2}{8G\ell^2}, \quad J=\frac{r_+ r_-}{4G\ell} .
\label{ab4}
\eeq
Kruskal coordinates are discussed in \cite{BHTZ}; the Penrose diagram
is essentially identical to that of an asymptotically anti-de Sitter
black hole in 3+1 dimensions.  Another useful coordinate system is
based on proper radial distance $\rho$ and two light-cone-like
coordinates $u,v = t/\ell\pm\phi$ \cite{Banados}; the metric then
takes the form
\beq
ds^2 = 4G\ell\left(L^+du^2 + L^-dv^2\right) - \ell^2d\rho^2
  + \left(\ell^2e^{2\rho} + 16G^2{L^+L^-}e^{-2\rho}\right)dudv
\label{ab5}
\eeq
with
\beq
L^\pm = \frac{(r_+ \pm r_-)^2}{16G\ell} .
\label{ab6}
\eeq
In these coordinates, the Chern-Simons connections (\ref{aa1}) take 
the simple form
\beq
A^{(+)} = \left( 
  \begin{array}{cc} \frac{1}{2}d\rho & -\frac{4G}{\ell}L^+e^{-\rho}du \\
  -e^{\rho}du & -\frac{1}{2}d\rho \end{array}\right) ,\quad
A^{(-)} = \left( 
  \begin{array}{cc} -\frac{1}{2}d\rho &  -e^{\rho}dv \\  
  -\frac{4G}{\ell}L^-e^{-\rho}dv & \frac{1}{2}d\rho \end{array}\right)
\label{ab7}
\eeq
It is easy to check that these connections satisfy the equations of
motion (\ref{aa4}).  This solution may be generalized: the Einstein 
field equations are still satisfied if one allows $L^+$ to be an arbitrary 
function of $u$ and $L^-$ to be an arbitrary function of $v$.  This 
dependence can be removed by a suitable diffeomorphism, but as we shall 
see later, such a diffeomorphism does not satisfy appropriate boundary
conditions at infinity.  $L^+$ and $L^-$ are thus examples of the
``would-be pure gauge'' degrees of freedom discussed above.

As a constant curvature spacetime, the BTZ black hole is locally isometric
to anti-de Sitter space.  In fact, it is globally a quotient space of
$\mathit{AdS}_3$ by a discrete group.  We can identify $\mathit{AdS}_3$
with the universal covering space of the group $\mathrm{SL}(2,\mathbb{R})$;
the BTZ black hole is then obtained by the identification \cite{BHTZ,Carlipb}
\beq
g \sim \rho^- g \rho^+, \qquad 
\rho^\pm = \left(\begin{array}{cc} e^{\pi(r_+ \pm r_-)/\ell} & 0 \\
       0 & e^{-\pi(r_+ \pm r_-)/\ell}\end{array}\right) .
\label{ab7a}
\eeq
Up to a gauge transformation, the group elements $\rho^\pm$ can be identified 
with the holonomies of the $\mathrm{SL}(2,\mathbb{R})$ connections (\ref{ab7}).

For our purposes, the most important feature of the BTZ black hole 
is that it has thermodynamic properties closely analogous to those 
of realistic (3+1)-dimensional black holes: it radiates at a Hawking 
temperature of
\beq
T = \frac{\hbar\kappa}{2\pi} = \frac{\hbar(r_+^2-r_-^2)}{2\pi\ell^2r_+} ,
\label{ab8}
\eeq
where $\kappa$ is the surface gravity, and has an entropy
\beq
S = \frac{2\pi r_+}{4\hbar G}
\label{ab9}
\eeq
equal to a quarter of its area.  These features can be obtained in
all the usual ways: from quantum field theory in a BTZ
background \cite{BTZ,Hyun,Ichi}; from Euclidean path integration 
\cite{CarTeit}; from the Brown-York microcanonical path integral
\cite{BMann}; from Wald's Noether charge approach \cite{Carlipf,CGeg};
and from tunneling arguments \cite{EngRez,Medved}.  There is even a powerful new 
method available \cite{Emparan}: one can consider quantum gravitational
perturbations induced by a classical scalar source, and then use detailed
balance arguments to obtain thermodynamic properties.  Together, these 
results strongly suggest that many of the mysteries of black hole 
statistical mechanics in higher dimensions can be investigated in 
this simpler setting as well.

\section{Gauge Invariances and Symmetries \label{GIS}}

I argued in section \ref{tog} that most of the degrees of freedom for
(2+1)-dimensional anti-de Sitter gravity represent excitations that would 
naively be considered ``pure gauge,'' but that become physical at the
conformal boundary.  As a first step in obtaining these degrees of freedom,
one must understand a rather subtle distinction between gauge invariances
and symmetries on manifolds with timelike boundaries.  The difference between
``proper'' and ``improper'' gauge transformations was first, I believe, 
studied in detail by Benguria et al.\ \cite{Benguria}, although it was 
to some extent implicit in \cite{RegTeit,DeWitt}.  This separation of
gauge transformations and symmetries has had some interesting applications 
in gauge theories \cite{Gervais}, but while the distinction is well 
known among experts, it is rarely explained clearly.

Let us begin with perhaps the simplest example \cite{Bal}.
Consider the Chern-Simons action (\ref{aa3}) on a manifold with the
topology $\mathbb{R}\times\Sigma$, where $\Sigma$ is a two-manifold with 
boundary $\partial\Sigma$.  In canonical form, the action becomes
\beq
I_{\mathit{CS}} = \frac{k}{4\pi}\int dt \int_\Sigma d^2x\,\epsilon^{ij}
  \mathrm{Tr}\left({\dot A}_iA_j + A_0F_{ij}\right) ,
\label{ca1}
\eeq
from which we can read off the Poisson brackets
\beq
\left\{A_i^a(x),A_j^b(x')\right\} 
  = \frac{2\pi}{k}\epsilon_{ij}{\hat g}^{ab}\delta^2(x-x')
\label{ca2}
\eeq
where ${\hat g}^{ab} = \mathrm{Tr}(T^aT^b)$ is the Cartan-Killing metric on 
the gauge group.  It is apparent from (\ref{ca1}) that $A_0$ is a Lagrange 
multiplier.  The corresponding first class constraints
\beq
G^{(0)}_a = \frac{k}{4\pi}{\hat g}_{ab}\epsilon^{ij}F^b_{ij}
\label{ca3}
\eeq
generate gauge transformations; that is, if $\Sigma$ is closed, the smeared 
generators 
\beq
G^{(0)}[\eta] = \int_\Sigma d^2x\, \eta^aG^{(0)}_a
\label{ca4}
\eeq
have brackets
\beq
\left\{ G^{(0)}[\eta],A^a_k\right\} = D_k\eta^a = \delta_\eta A^a_k ,
\label{ca4a}
\eeq
where $D_k$ is the gauge-covariant derivative.  Further, the generators satisfy 
a Poisson algebra isomorphic to the gauge algebra,
\beq
\left\{ G^{(0)}[\eta],G^{(0)}[\xi]\right\} = G^{(0)}[\zeta], \quad
\zeta^c = f^c{}_{ab}\eta^a\xi^b ,
\label{ca5}
\eeq
where $f^c{}_{ab}$ are the structure constants of the gauge group.

We now make the crucial observation that on a manifold with boundary,
the generators $G_a$ are not ``differentiable'' \cite{RegTeit}: that is,
the functional derivative of a smeared generator $G^{(0)}[\eta]$ involves an
ill-defined surface term.  Indeed, a simple calculation shows that
\beq
\delta G^{(0)}[\eta] 
  = \frac{k}{2\pi}\int_\Sigma d^2x\,\epsilon^{ij}\eta_aD_i\delta A^a_j
  = - \frac{k}{2\pi}\int_\Sigma d^2x\,\epsilon^{ij}D_i\eta_a \delta A^a_j
    + \frac{k}{2\pi}\int_{\partial\Sigma}\eta_a \delta A^a_k dx^k ,
\label{ca6}
\eeq
and if $\eta\ne0$ on the boundary, the last term ruins the Poisson
algebra (\ref{ca5}).  To restore the algebra, one must add a boundary
term $Q[\eta]$ to $G^{(0)}[\eta]$, with a variation
\beq
\delta Q[\eta] = -\frac{k}{2\pi}\int_{\partial\Sigma}\eta_a \delta A^a_k dx^k .
\label{ca7}
\eeq
The full generator $G[\eta] = G^{(0)}[\eta] + Q[\eta]$ then has a well-defined
functional derivative, and a straightforward computation \cite{Banadosb} yields
the Poisson algebra 
\beq
\left\{ G[\eta],G[\xi]\right\} = G[\zeta(\eta,\xi)] 
  + \frac{k}{2\pi}\int_{\partial\Sigma}\eta_ad\xi^a .
\label{ca8}
\eeq
I have assumed here that the gauge parameters $\eta$ and $\xi$ are independent
of the fields, and therefore have vanishing Poisson brackets with the generators
$G$; see \cite{Banadosb,BanOrt} for a useful generalization to field-dependent 
parameters.

Equation (\ref{ca8}) can be recognized as a central extension of the original
algebra of gauge transformations.  We shall return to this point below.  But 
let us first consider the implications for the symmetries of our Chern-Simons
theory.

The quantity $G^{(0)}[\eta]$ vanishes by virtue of the field equations, and
its Poisson bracket with any physical observable $\cal O$ must therefore also
vanish: $\{ G^{(0)}[\eta], {\cal O}\}=0$.  In the quantum theory, the Poisson
brackets become commutators, and the corresponding statement is that matrix 
elements of $[G^{(0)}[\eta], {\cal O}]$ between physical states must vanish.
On a compact manifold, this is simply the statement that physical observables 
must be gauge-invariant, and that objects that are ``pure gauge'' can have no 
physical meaning.  Similarly, the vanishing of $G^{(0)}[\eta]$ normally implies 
that
\beq
G^{(0)}[\eta] |\mathit{phys}\rangle = 0 ,
\label{ca8a}
\eeq
although this condition can be weakened slightly, for example by requiring it
to hold only for the positive frequency components of $G^{(0)}[\eta]$.

If $\Sigma$ has a boundary, on the other hand, the generator of gauge transformations 
is not $G^{(0)}[\eta]$, but rather $G[\eta]$.  In general, the boundary contribution
$Q[\eta]$ to $G[\eta]$ need not vanish.  Indeed, if $\eta\ne0$ at $\partial\Sigma$,
setting $G[\eta]$ to zero would be inconsistent with (\ref{ca8}).  Hence physical
observables need not be invariant under gauge transformations at the boundary;
it is enough that they transform under some representation of the algebra (\ref{ca8}).
Equivalently, if we look at the algebra of the original constraints $G^{(0)}[\eta]$ 
in the presence of a boundary, we find \cite{Parkb,Fjel}
\beq
\left\{ G^{(0)}[\eta],G^{(0)}[\xi]\right\} = G^{(0)}[f^c{}_{ab}\eta^b\xi^c] 
  + \hbox{\it $\delta$-function boundary terms} .
\label{ca9}
\eeq
This means that the constraints $G^{(0)}[\eta]$ become second class at the
boundary, and we can no longer consistently impose condition (\ref{ca8a}).

Gauge transformations are thus very different in the bulk and at a boundary: in 
the bulk they are true invariances, but at a boundary they are only symmetries.  
While I have shown this in detail for Chern-Simons theory, a similar argument 
holds for any gauge or gauge-like theory.  This, as we shall see, has very
important implications for the physics of (2+1)-dimensional gravity.

\section{Asymptotic Symmetries, AdS/CFT, and State-Counting}
\setcounter{footnote}{0}

The first hint that boundary degrees of freedom can account for the
entropy of the BTZ black hole comes from a symmetry argument that
does not require knowledge of the detailed dynamics \cite{Stromb,BSS}.  
The boundary of an asymptotically anti-de Sitter space is a cylinder, 
both topologically and metrically, and it is not surprising that the
asymptotic diffeomorphisms are related to the diffeomorphisms of
a cylinder.  What is somewhat surprising is that the resulting
Virasoro algebra contains a central term \cite{BrownHenn}, which 
can be determined by standard methods within the
framework of classical general relativity.  One can then appeal
to the remarkable result, due to Cardy \cite{Cardy,Cardyb}, that 
the asymptotic density of states in a two-dimensional conformal field 
theory is fixed by a few features of the symmetry algebra, independent
of any details of the dynamics.\footnote{As far as I know, the idea
of using the Cardy formula to count BTZ black hole states was first
suggested in \cite{CarTeit}, but the authors of that paper were unaware
of the computation of the central charge in \cite{BrownHenn}, and
therefore gave only qualitative arguments.}  While such a symmetry argument 
does not explain the underlying quantum mechanical degrees of freedom, it
does suggest a solution to the problem of universality \cite{Carlipa}
discussed in the Introduction. 

\subsection{Asymptotic Symmetries \label{AS}}

The asymptotic symmetries of (2+1)-dimensional asymptotically anti-de Sitter 
space were first investigated by Brown and Henneaux \cite{BrownHenn,Brown}.
The analysis involves two key steps.  First, one must find the diffeomorphisms 
that preserve the asymptotic structure of the anti-de Sitter metric (\ref{ab1}).  
It is straightforward to show that these are generated by vector fields of 
the form
\begin{alignat}{2}
\xi^{(+)t} &= \ell T^+ + \frac{\ell^3}{2r^2}\partial_u^2T^+
  + \mathcal{O}\left(\frac{1}{r^4}\right) &\qquad
\xi^{(-)t} &= \ell T^- + \frac{\ell^3}{2r^2}\partial_v^2T^- 
  + \mathcal{O}\left(\frac{1}{r^4}\right)
\nonumber\\
\xi^{(+)\phi} &= \ell T^+ - \frac{\ell^3}{2r^2}\partial_u^2T^+ 
  + \mathcal{O}\left(\frac{1}{r^4}\right)&\qquad
\xi^{(-)\phi} &= -\ell T^- + \frac{\ell^3}{2r^2}\partial_v^2T^- 
  + \mathcal{O}\left(\frac{1}{r^4}\right)\label{ba1}\\
\xi^{(+)r} &= -r\partial_uT^+ + \mathcal{O}\left(\frac{1}{r}\right)&\qquad
\xi^{(-)r} &= -r\partial_vT^- + \mathcal{O}\left(\frac{1}{r}\right) \nonumber
\end{alignat}
where, as before, $u,v = t/\ell\pm\phi$ and $T^\pm$ are functions of $u$ 
and $v$, respectively.  It may be checked that the algebra of such vector
fields is closed under commutation: $\xi^{(+)}$ and $\xi^{(-)}$ commute, 
while the commutators $[\xi_1^{(\pm)},\xi_2^{(\pm)}] = \xi_{[1,2]}^{(\pm)}$
yield new vectors of the form (\ref{ba1}) with
\begin{align}
T_{[1,2]}^+ &= 2\left(T_1^+\partial_uT_2^+ - T_2^+\partial_uT_1^+\right),
\nonumber\\
T_{[1,2]}^- &= 2\left(T_1^-\partial_vT_2^- - T_2^-\partial_vT_1^-\right).
\label{ba2}
\end{align}
Eqn.\ (\ref{ba2}) may be recognized as a pair of Virasoro algebras, each with
vanishing central charge \cite{CFT}.

The second step is to realize these symmetries as canonical transformations.
As in (3+1)-dimensional gravity, the gauge transformations are generated by
the Hamiltonian and momentum constraints, whose algebra is (up to some
subtleties \cite{Ryan}) the algebra of diffeomorphisms.  For a noncompact
manifold, however, there is an added complication: as in section \ref{GIS},
boundary terms must be added to the constraints to make them differentiable 
\cite{RegTeit}.  The canonical generators thus take the general form
\beq
H[\xi] = \int d^2x\,\xi^\mu\mathcal{H}_\mu + J[\xi]
\label{ba3}
\eeq
where the boundary terms $J[\xi]$ are chosen in such a way that the
functional derivatives $\delta H/\delta g_{ab}$ and $\delta H/\delta\pi^{ab}$,
and thus the Poisson brackets of the generators, are well-defined.  The
presence of such boundary terms can alter the Poisson brackets: in general,
a central term appears \cite{BrownHenn,BrownHennb},
\beq
\left\{H[\xi],H[\eta]\right\} = H[\{\xi,\eta\}] + K[\xi,\eta] .
\label{ba4}
\eeq
For asymptotically anti-de Sitter space, and in particular for the BTZ black 
hole, the net effect is that the Virasoro algebras (\ref{ba2}) acquire a central 
charge:  
\begin{align}
\left\{L^\pm_m,L^\pm_n\right\} &= i(m-n)L^\pm_{m+n} + \frac{ic}{12}m(m^2-1)\delta_{m+n,0} 
  \nonumber\\
\left\{L^+_m,L^-_n\right\} &= 0 ,\label{ba5} 
\end{align}
where $L^\pm_n=H[\xi^{(\pm)}_n]$ are the canonical generators of the 
asymptotic diffeomorphisms (\ref{ba1}) with $T^+_n=e^{-inu}$ and 
$T^-_n=e^{-inv}$, and where
\beq
c = \frac{3\ell}{2G}
\label{ba6}
\eeq
is the central charge.  The asymptotic conserved charges $L_0^\pm$ 
coming from constant $T^\pm$ are linear combinations of the mass 
(associated with a constant time translation $\xi^t$) and angular
momentum (associated with a constant rotation $\xi^\phi$), and a 
direct computation yields\footnote{The Brown-Henneaux paper \cite{BrownHenn} 
actually appeared before the discovery of the BTZ black hole, and discussed 
conical singularities in asymptotically anti-de Sitter space rather than 
black holes.  But the boundary conditions were deliberately chosen to accommodate
more general solutions, and the generalization to the BTZ metric is immediate.}
\beq
L^\pm = \frac{(r_+ \pm r_-)^2}{16G\ell} ,
\label{ba7}
\eeq
in agreement with the notation of (\ref{ab6}).

Terashima has shown that the same central charge can be obtained from a path 
integral by way of the Ward-Takahashi identities \cite{Terashima,Terashimab}.
A similar, and in some ways simpler, derivation also exists in the Chern-Simons
formalism \cite{Banadosb,Oh}.  Recall from (\ref{aa5}) that a diffeomorphism
is represented in Chern-Simons theory as a gauge transformation with parameter
$\eta^a = \xi^\mu A^a_\mu$.  From (\ref{ca7}), the corresponding boundary term 
in the generator of gauge transformations is
\beq
\delta Q[\eta] 
  = -\frac{k}{2\pi}\int_{\partial\Sigma}\left(
  \xi^\rho {\hat g}_{ab}A^a_\rho\delta A^b_\phi 
  + \xi^\phi {\hat g}_{ab}A^a_\phi\delta A^b_\phi
  \right) d\phi .
\label{ba7a}
\eeq
From (\ref{ba1}), $\xi^{(\pm)\rho} = -\frac{1}{\ell}\partial_\phi\xi^{(\pm)\phi}$,
while from (\ref{ab7}), $A_\rho{}^{(\pm)} = \alpha^{(\pm)}$ should be fixed at 
infinity.  We can thus integrate (\ref{ba7a}) to obtain
\beq
Q^{(\pm)}[\xi] = -\frac{k}{4\pi}\int\left(
  \xi^{(\pm)\phi} {\hat g}_{ab}A_\phi^{(\pm)a} A_\phi^{(\pm)b}
  + 2\xi^{(\pm)\phi} {\hat g}_{ab}\alpha^{(\pm)a} \partial_\phi A_\phi^{(\pm)b} 
  + \xi^{(\pm)\phi} {\hat g}_{ab}\alpha^{(\pm)a}\alpha^{(\pm)b} \right) d\phi ,
\label{ba8}
\eeq
where the last term is an integration constant.  Given these generators and 
the Poisson brackets (\ref{ca2}), it is straightforward to verify the algebra 
(\ref{ba5}) with central charges
\beq
c^{(\pm)} = \frac{3\ell}{G}\alpha^{(\pm)}_a\alpha^{(\pm)a} .
\label{ba9}
\eeq 
In particular, for the connections (\ref{ab7}), one can read off the
values $\alpha^{(\pm)}_a\alpha^{(\pm)a} = 1/2$; (\ref{ba9}) then  
reproduces the central charge (\ref{ba6}).  As discussed in section \ref{WZWL}, 
the Chern-Simons formulation can be further reduced to Liouville theory
\cite{Coussaert}; the resulting central charge again agrees with 
(\ref{ba6}).

A further confirmation of this value of the central charge comes from 
considering the symmetries of the metric (\ref{ab5}), or, more simply, 
the Chern-Simons connections (\ref{ab7}).  The most general infinitesimal 
transformation that preserves the form of $A^{(+)}$ is parametrized by a 
function $\epsilon(u)$, and one finds that \cite{Banados,Ezawab,Navarro}
\beq
\delta L^+ = \epsilon\partial_uL^+ + 2(\partial_u\epsilon) L^+ +
  \frac{\ell}{8G}\partial_u^3\epsilon .
\label{ba10}
\eeq
This may be recognized as the transformation law for the holomorphic part
of a stress-energy tensor in a conformal field theory with central charge
(\ref{ba6}) \cite{CFT}.  This derivation has been extended to supergravity 
in \cite{BanBaut}, where it is shown that one can obtain a super-Virasoro 
algebra.  Note that while the transformation (\ref{ba10}) respects our 
asymptotically anti-de Sitter boundary conditions, it certainly acts 
nontrivially on boundary values of the fields.  As such, it must be considered
a symmetry rather than a gauge transformation; that is, as discussed in 
section \ref{GIS}, the asymptotic fields should transform under some 
representation of the algebra (\ref{ba5}), but need not be invariant.   

Yet another derivation of the central charge (\ref{ba6}) comes from
considering the quasilocal stress-energy tensor at the conformal boundary
of asymptotically anti-de Sitter space \cite{BalKraus}.  For a region 
$U$ with timelike boundary $\partial U$, with induced metric $\gamma_{ij}$ on
$\partial U$, the Brown-York quasilocal stress-energy tensor \cite{BrownYork} 
is defined as
\beq
T^{ij} = \frac{2}{\sqrt{|\gamma|}}
  \frac{\delta I_{\mathit{EH}}}{\delta \gamma_{ij}}
\label{ba11}
\eeq
where $I_{\mathit{EH}}$ is the Einstein-Hilbert action for $U$ with an 
appropriate boundary term added to ensure that the variational principle 
is well defined for fixed $\gamma_{ij}$ \cite{Gibbons,Wald}: in three 
dimensions,
\beq
I_{\mathit{EH}} = \frac{1}{16\pi G}\int_U d^3x\left( R + \frac{2}{\ell^2}\right)
  + \frac{1}{8\pi G}\int_{\partial U}d^2x\sqrt{|\gamma|}K ,
\label{ba12}
\eeq
where $K$ is the extrinsic curvature of $\partial U$.  For asymptotically
AdS spacetimes, though, the stress-energy tensor (\ref{ba11}) diverges as
the boundary approaches conformal infinity.  This divergence may be cured by 
adding a local counterterm $I_{\mathit{ct}}$ \cite{BalKraus,HennSken}; in 2+1 
dimensions the appropriate term is
\beq
I_{\mathit{ct}} = - \frac{1}{8\pi G\ell}\int_{\partial U}d^2x\sqrt{|\gamma|} .
\label{ba13}
\eeq
With this choice, the trace of the Brown-York stress-energy tensor becomes
\beq
T = -\frac{\ell}{16\pi G}{}^{(2)}R ,
\label{ba14}
\eeq
where ${}^{(2)}R$ is the curvature scalar for the boundary metric $\gamma_{ij}$.
But in a general two-dimensional conformal field theory, the conformal anomaly
is \cite{CFT}
\beq
T = -\frac{c}{24\pi}{}^{(2)}R ,
\label{ba15}
\eeq
agreeing with (\ref{ba14}) if $c=3\ell/2G$.  This derivation is closely related
to the Fefferman-Graham construction of Liouville theory that will be discussed
below in section \ref{FG}; indeed, as we shall see, the anomaly (\ref{ba14})
can be computed directly from an expansion of the metric and action near
infinity \cite{HennSken}.

Relationships among some of these results have been discussed in \cite{BanOrtb}.
While the derivation of the Virasoro algebra (\ref{ba5}) and central charge 
(\ref{ba6}) is evidently quite robust, it is worth pointing out that the
value of the central charge depends on the choice of boundary conditions.
As noted in \cite{Carlipg}, for example, a boundary condition that fixes 
$A_\rho$ at some finite radius $r$ rather than at infinity will lead to a
``blue-shifted'' central charge.  I shall return to this issue in section
\ref{what}.

\subsection{The Cardy Formula \label{CF}}

When the central charge in the symmetry algebra of (2+1)-dimensional 
asymptotically AdS gravity was first discovered, it was considered to be
mainly a mathematical curiosity.  This changed when Strominger \cite{Stromb}
and Birmingham, Sachs, and Sen \cite{BSS} independently pointed out that
this result could be used to compute the asymptotic density of states.
The key to this computation is the Cardy formula \cite{Cardy,Cardyb}.

Consider a two-dimensional conformal field theory, whose symmetries are
described by a Virasoro algebra with central charge $c$.  Let $\Delta_0$
be the smallest eigenvalue of $L_0$ in the spectrum, and define an effective
central charge 
\beq
c_{\hbox{\scriptsize\it eff}} = c-24\Delta_0 .
\label{bb1}
\eeq
Then for large $\Delta$, the density of states with eigenvalue $\Delta$
of $L_0$ is\footnote{In conformal field theory parlance, $\Delta$ is a
``conformal weight.''}
\beq
\rho(\Delta) \approx
\exp\left\{2\pi\sqrt{\frac{c_{\hbox{\scriptsize\it eff}}\Delta}{6}}\right\}
\rho(\Delta_0) .
\label{bb2}
\eeq
A careful proof of this result using the method of steepest descents is
given in \cite{Carlipg}.  One can derive the logarithmic corrections to 
the entropy by the same methods \cite{Carliph}; and indeed, by using results 
from the theory of modular forms, one can obtain even higher order corrections 
\cite{Birm,Farey}.

Although the mathematical derivation of the Cardy formula is relatively 
straightforward, I do not know of a good, intuitive \emph{physical}
explanation for (\ref{bb2}).  The derivation relies on a duality between
high and low temperatures, which arises from modular invariance: by
interchanging cycles on a torus, one can trade a system on a circle
of circumference $L$ with inverse temperature $\beta$ for a system on
a circle of circumference $\beta$ with inverse temperature $L$.  But it
would be helpful to have a more direct understanding of why the density
of states is identical for systems with very different physical degrees
of freedom, but with the same values of $c$ and $\Delta_0$.

For a few cases, the behavior (\ref{bb2}) \emph{does} have a more immediate
explanation.  A free scalar field, for example, has creation operators 
$a_{-n}$, with $[L_0,a_{-n}] = na_{-n}$.  If we choose a standard vacuum,
for which $a_n |0\rangle = 0$ for $n\ge0$, then excited states created by 
applying a chain of creation operators will satisfy
\beq
L_0 \left( a_{-n_1}a_{-n_2}\dots a_{-n_m}\right) |0\rangle  =
(n_1 + n_2 + \dots + n_m) \left( a_{-n_1}a_{-n_2}\dots a_{-n_m}\right)|0\rangle .
\label{bb3}
\eeq
The number of states for which $L_0$ has eigenvalue $\Delta$ is thus simply 
the number of distinct ways of writing $\Delta$ as a sum of integers.  This 
is the famous partition function $p(\Delta)$ of number theory, whose asymptotic 
behavior is \cite{Ramanujan}
\beq
\ln p(\Delta) \sim 2\pi\sqrt{\Delta/6} \ ,
\label{bb4}
\eeq
matching the prediction of the the Cardy formula for $c=1$.  More generally,
combinatoric methods can be applied if we start with a set of bosonic ``creation 
operators'' $\phi^{(M_n)}_n$, with conformal dimensions
\beq
[L_0,\phi^{(M_n)}_n] = \beta n\phi^{(M_n)}_n ,
\label{bb5}
\eeq
where $\beta$ is a constant and the index $M_n$ distinguishes fields
with identical dimensions.  Let $\gamma(n)$ denote the degeneracy at
conformal dimension $\beta n$, i.e., $M_n=1,\dots,\gamma(n)$.  We allow
$\gamma(n)$ to be zero for some values of $n$---the conformal dimensions
need not be equally spaced.  Then if the asymptotic behavior of the sum of 
degeneracies is of the form
\beq
\sum_{n\le x}\gamma(n) \sim Kx^u 
\label{bb6}
\eeq
for large $x$, it can be shown that the number of states with $L_0 = \Delta$ 
grows as \cite{Brigham}
\beq
\ln \rho(\Delta) \sim \frac{1}{u}[u+1]^{u/(u+1)}
\left[ Ku\Gamma(u+2)\zeta(u+1)\right]^{1/(u+1)}
\left[ {\Delta/\beta} \right]^{u/(u+1)} .
\label{bb7}
\eeq
 
We can now apply the Cardy formula to the BTZ black hole.  To do so, we shall
assume that $\Delta_0=0$, so $c_{\hbox{\scriptsize\it eff}}$ is given, up to
quantum corrections, by (\ref{ba6}).  This assumption can certainly be 
questioned \cite{Martinecb}, and I shall return to it in section \ref{count}.
Given such a central charge, though, the Cardy formula for the classical
charges (\ref{ba7}) yields and entropy
\beq
S = \ln\rho(\Delta^+) + \ln\rho(\Delta^-) 
  = 2\pi\left(\sqrt{\frac{cL^+}{6}} + \sqrt{\frac{cL^-}{6}}\right)
  = \frac{2\pi r_+}{4G} ,
\label{bb8}
\eeq
agreeing precisely with the Bekenstein-Hawking entropy (\ref{ab9}).

Note that the key to this result is the existence of a \emph{classical} 
central charge.  Indeed, if one is careful about factors of $\hbar$,
the classical Poisson brackets $\{L,L\}\sim L + c$ become quantum
commutators $[L/\hbar,L/\hbar]\sim L/\hbar + c/\hbar$, giving the factor
of $\hbar$ in the usual Bekenstein-Hawking entropy (see, for instance,
\cite{Parkc}). A quantum mechanical central charge, on the other hand, 
will typically be of order $1$, as will a quantum correction to the 
classical value of $L_0$.  Thus if one wishes to obtain $c$ and $L_0$ 
strictly as quantum corrections to a symmetry algebra with no classical 
central charge, one must generate enormously large values of $c$ and 
$L_0$.  This may be possible in certain models---in Sakharov-style 
induced gravity, for instance, one can obtain an effective Liouville 
theory whose ``classical'' central charge is entirely due to quantum 
effects for a very large number of heavy constituent fields \cite{Frolovb}.

While the state-counting argument described here was originally applied 
to the BTZ black hole, the same method correctly counts states in a variety 
of other asymptotically anti-de Sitter solutions, including black holes
coupled to scalar fields \cite{Natsuume,Natsuumeb,Park} and black holes
in gravitational theories with higher-order curvature terms \cite{Saida}.  
On the other hand, since the method depends only on the asymptotic behavior 
of the metric, it also gives an entropy (\ref{bb8}) for a ``star'' \cite{Cruz,Lubo}, 
a circularly symmetric, horizonless lump of matter whose exterior is described 
by the BTZ metric.  I shall return to this issue in section \ref{what}.

\subsection{The Effective Central Charge \label{ECC}}

It will be important later that the central charge occurring in the Cardy
formula is the effective central charge $c_{\hbox{\scriptsize\it eff}}$.
Before exploring the significance of this fact, a few subtleties in notation
need to be clarified.

The Virasoro algebra (\ref{ba5}) is the algebra of holomorphic
diffeomorphisms of a conformal field theory on the complex plane, with
\beq
L_n = i\int dz\,z^{n+1}T_{zz}
\label{bc1}
\eeq
in the conventions of \cite{CFT}.  One can transform to the cylinder with a 
mapping $z=e^{r+i\phi}$, and define generators $L_n^{{\hbox{\scriptsize\it cyl}}}$
as the Fourier components of $T_{\phi\phi}$; because of the anomalous 
transformation properties of the stress-energy tensor, one finds that
\beq
L_n^{{\hbox{\scriptsize\it cyl}}} = L_n - \frac{c}{24}\delta_{n0} ,
\label{bc2}
\eeq
and the algebra (\ref{ba5}) becomes
\beq
[L_m^{{\hbox{\scriptsize\it cyl}}},L_n^{{\hbox{\scriptsize\it cyl}}}] 
  = (m-n)L_{m+n}^{{\hbox{\scriptsize\it cyl}}} + \frac{c}{12}m^3\delta_{m+n,0} .
\label{bc3}
\eeq
The Virasoro eigenvalue $\Delta_0$ appearing in the definition of
$c_{\hbox{\scriptsize\it eff}}$ is the lowest eigenvalue of $\Delta_0$
on the plane; the corresponding eigenvalue (\ref{bc2}) on the cylinder is
shifted by the Casimir energy of the fields on a compact space.

For a simple illustration of an ``effective central charge,''
now consider a standard affine Lie algebra\footnote{This algebra has
appeared in equation (\ref{ca8}) as the algebra of constraints of a
Chern-Simons theory on a manifold with boundary.}
\beq
J^a = \sum_n J^a_n e^{in\phi}, \qquad
[J^a_m,J^b_n] = if^{ab}{}_cJ^c_{m+n} - km {\hat g}^{ab}\delta_{m+n,0} ,
\label{da1}
\eeq
with corresponding Virasoro generators given by the Sugawara construction
\cite{CFT},
\beq
L_n = \frac{1}{2(k+h)}\sum_{m=-\infty}^{\infty} {\hat g}_{ab}:J^a_mJ^b_{n-m}:
\label{da2}
\eeq
where, as in section \ref{AS}, ${\hat g}_{ab}$ is the Cartan-Killing metric, 
and $h$ is the dual Coxeter number of $G$,
\beq
f^{ab}{}_cf^{dc}{}_b = -2h{\hat g}^{ad} .
\label{da2a}
\eeq
It is straightforward to check that these $L_n$ satisfy the algebra (\ref{ba5}),
with a central charge determined by the group, and that the asymptotic density
of states is given by the Cardy formula.  Now, however, consider the deformed 
Virasoro algebra \cite{Freericks,Sakai} generated by
\beq
{\tilde L}_n
  = L_n + in\alpha_aJ^a_n + \frac{k}{2}\alpha_a\alpha^a\delta_{n0} .
\label{bc4}
\eeq
It is easy to check that the ${\tilde L}_n$ again satisfy the Virasoro
algebra (\ref{ba5}), but with a new central charge
\beq
{\tilde c} = c + 12k\alpha_a\alpha^a .
\label{bc5}
\eeq
But the redefinition (\ref{bc4}) has not changed the Hilbert space, so
the asymptotic behavior of the density of states should not be affected.

In fact, it is not.  Under the deformation (\ref{bc4}), $L_0$ has shifted
by a constant, and its lowest eigenvalue is now
\beq
{\tilde\Delta}_0 = \Delta_0 + \frac{k}{2}\alpha_a\alpha^a .
\label{bc6}
\eeq
The effective central charge, and hence the density of states predicted
by the Cardy formula, is thus invariant.

We shall see in section \ref{CSWZW} that the boundary value $A^a_\mu$ of
a Chern-Simons gauge field can be identified with an affine current having
an algebra of the form (\ref{da1}).  As a consequence, the Virasoro generators 
(\ref{ba8}) can be understood as deformed generators of precisely the form
(\ref{bc4}).  This suggests that we should treat the Cardy formula derivation
of the BTZ black hole entropy with caution \cite{Martinecb}---it
is not at all obvious that the classical central charge of section \ref{AS}
is the correct effective central charge.  On the other hand, if the effective
central charge is substantially different from the classical value (\ref{ba6}),
the success of the counting arguments of the preceding section would become
an extraordinary coincidence, crying out for a deeper explanation.

\section{Reducing Gravity to Conformal Field Theory \label{GCFT}}
\setcounter{footnote}{0}

The analysis of the preceding section suggests that quantum general relativity
can give the correct counting of microscopic states needed to explain the
entropy of the BTZ black hole.  It does not, however, tell us what those
states are.  The main virtue of the Cardy formula, its indifference to the
details of the states being counted, is also its main weakness---we can
count states without a full quantum theory of gravity, but the actual states
remain disguised.  

Whether the states of the BTZ black hole can be obtained purely within the
framework of gravity has been a hotly debated question.  Martinec, for
example, has argued that general relativity must be considered an effective
field theory, which cannot distinguish among different conformal field
theory states with the same expectation values of the stress-energy tensor; 
only a more complete microscopic theory (string theory or a dual gauge
theory) can describe the true underlying degrees of freedom \cite{Martinecb}.  
On the other hand, a number of authors have attempted---with varying degrees 
of success---to obtain the BTZ black hole entropy by counting states in 
particular conformal field theories that are, arguably, induced from pure 
(2+1)-dimensional gravity.  To evaluate these attempts, one must first 
understand how to obtain such conformal field theories.  

\subsection{From Chern-Simons to Wess-Zumino-Witten \label{CSWZW}}

The equations of motion for a Chern-Simons theory are that the field strength
$F_{\mu\nu}$ vanishes.  On a topologically trivial manifold---say, $M =
\mathbb{R}\times D^2$---this implies that the potential $A$ is ``pure gauge,''
\beq
A = g^{-1}dg .
\label{cb1}
\eeq
As one might expect from section \ref{GIS}, though, the gauge parameter $g$
can have nontrivial dynamics on the boundary.

As Witten first suggested \cite{Wittena}, this dynamics can be described by
a Wess-Zumino-[Novikov]-Witten (WZ[N]W) model \cite{Wittenc}.  Perhaps the simplest 
way to understand this relation \cite{MS,EMSS} is to begin with the canonical 
formalism of section \ref{GIS} and substitute $A_i = g^{-1}\partial_ig$ into 
the action (\ref{ca1}).  A straightforward computation shows that the resulting 
action for $g$ is a WZW action on the boundary $\partial M = \mathbb{R}\times S^1$.  

A closely related but slightly more general approach \cite{Ogura,Carlipi} starts
with the observation \cite{Jackiw} that on a general three-manifold with boundary,
the Chern-Simons action is not quite gauge invariant: under a gauge transformation
\beq
A = g^{-1}dg + g^{-1}{\bar A}g ,
\label{cb2}
\eeq
the action (\ref{aa3}) transforms as 
\beq
I_{\mathit{CS}}[A] =  I_{\mathit{CS}}[{\bar A}] 
  - \frac{k}{4\pi}\int_{\partial M}\mathrm{Tr}\left((dgg^{-1})\wedge{\bar A}\right)
  - \frac{k}{12\pi}\int_M\mathrm{Tr}\left(g^{-1}dg\right)^3 .
\label{cb3}
\eeq
For a closed manifold, the last term in (\ref{cb3}) is proportional to a topological 
invariant, a winding number \cite{Wittenc}; for $k$ an integer, $I_{\mathit{CS}}$ 
shifts by $2\pi k N$, so despite first appearances, $\exp\{iI_{\mathit{CS}}\}$ is 
invariant.  For a manifold with boundary, however, this term cannot, in general,
be discarded.

A further gauge dependence appears because one must add a surface term to the 
action when $M$ is not compact.  For a manifold with boundary, the Chern-Simons 
action has no extrema: a variation of $A$ gives
\beq
\delta I_{\mathit{CS}}[A] 
  = \frac{k}{2\pi}\int_M \mathrm{Tr}\left[\delta A\left(dA + A\wedge A\right)\right]
            -\frac{k}{4\pi}\int_{\partial M} \mathrm{Tr}\left(A\wedge\delta A\right) ,
\label{cb4}
\eeq
and as in section \ref{GIS}, one must add a boundary contribution to the action 
to cancel the boundary term in (\ref{cb4}).  The form of this new term will depend 
on our choice of boundary conditions.  Good boundary conditions generally require 
that we fix half the canonical data---positions but not momenta, for instance---but 
from the Poisson brackets (\ref{ca2}), the gauge potentials $A$ are \emph{both} 
positions \emph{and} momenta.  We thus need additional information to separate out 
``half the data'' to be held constant at $\partial M$.

Typically, this added information comes in the form of a choice of complex structure 
on $\partial M$.  If we choose such a complex structure and prescribe a fixed boundary 
value for, say, $A_z$, the appropriate boundary term in the action is easily seen to be
\beq
I_{\mathit{bdry}}[A] = \frac{k}{4\pi}\int_{\partial M}\mathrm{Tr} A_zA_{\bar z}, 
\label{cb5}
\eeq
which transforms as
\beq
I_{\mathit{bdry}}[A] = I_{\mathit{bdry}}[{\bar A}] + \frac{k}{4\pi}\int_{\partial M} 
  \mathrm{Tr}\left(\partial_z g\,g^{-1}\partial_{\bar z}g\,g^{-1}
  + \partial_z g\,g^{-1}{\bar A}_{\bar z} + \partial_{\bar z}g\,g^{-1}{\bar A}_z \right) .
\label{cb6}
\eeq
Combining (\ref{cb3}) and (\ref{cb6}), we see that
\beq
(I_{\mathit{CS}}+I_{\mathit{bdry}})[A] = (I_{\mathit{CS}}+I_{\mathit{bdry}})[{\bar A}]
  + k I^+_{\mathit{WZW}}[g^{-1},{\bar A}] ,
\label{cb7}
\eeq
where
\beq
I^+ _{\mathit{WZW}}[g^{-1},{\bar A}_z]
 = \frac{1}{4\pi}\int_{\partial M}\mathrm{Tr}
 \left(g^{-1}\partial_z g\,g^{-1}\partial_{\bar z}g
 - 2g^{-1}\partial_{\bar z}g {\bar A}_z\right)
 + \frac{1}{12\pi}\int_M\mathrm{Tr}\left(g^{-1}dg\right)^3
\label{cb8}
\eeq
is the chiral WZW action for $g$ coupled to a background field ${\bar A}_z$.  

As anticipated, the gauge parameter $g$ has become dynamical at the boundary 
$\partial M$.  In particular, in a path integral evaluation of the partition
function or correlators, one can perform the usual Faddeev-Popov trick of 
splitting the integral into an integral over $\bar A$ and one over $g$, but
since the action depends on $g$, the latter may no longer simply be divided 
out.  A similar phenomenon occurs in anomalous gauge theories \cite{Harada,%
Babelon}; the difference here is that the extra $g$-dependent piece appears
only at the boundary.

Note that the WZW current 
\beq
J_z = -k\partial_zg g^{-1} = -kgA_zg^{-1} + k{\bar A}_z 
\label{cb8a}
\eeq
is essentially the same as the gauge field $A_z$.  From the perspective of conformal 
field theory, the role of the background field ${\bar A}_z$ is to permit this
current to have a fixed, nontrivial holonomy.  For the BTZ black hole, in particular,
this holonomy is given by (\ref{ab7a}).  In some references, the background field is 
absorbed into $J_z$ by redefining $g$; but if the holonomy is nontrivial, such a 
redefinition requires that $g$ be multivalued.

In a related derivation of the WZW action, Fjelstad and Hwang start with the 
Chern-Simons action in the canonical form of section \ref{GIS} and note that the 
boundary term in (\ref{ca6}) makes the constraints second class at the boundary
\cite{Fjel}.  There is a standard procedure for restoring the full symmetry to 
a system with second class constraints: one can add new degrees of freedom that 
convert the second class constraints to first class constraints \cite{Faddeev}.  
Reference \cite{Fjel} shows that the new degrees of freedom are precisely those of 
a WZW theory, and that different gauge choices lead either to a pure chiral WZW 
theory or a pure Chern-Simons theory.  This derivation is essentially the converse 
of the one described above; instead of isolating ``would-be gauge'' degrees of 
freedom $g$, Fjelstad and Hwang show that one can add in new degrees of freedom 
$g$ to restore full invariance of the action.

A key test of these results comes from looking at the ``gluing'' of Chern-Simons 
theories.  As described above in section \ref{tog}, one can begin 
with a compact manifold $M$ and split it along a surface $\Sigma$ to obtain two 
manifolds $M_1$ and $M_2$, each with a boundary diffeomorphic to $\Sigma$.  
In general, the partition functions for $M_1$ and $M_2$ will depend on the boundary 
values of $A$, and one should be able to recover the partition function for $M$ by 
integrating over these values.  But as Witten has shown \cite{Wittenb}, one obtains
the correct composition law (\ref{a1}) only by including chiral WZW actions in the 
partition functions $Z_{(M_1,\Sigma)}$ and $Z_{(M_2,\Sigma)}$ from the start.

\subsection{From Wess-Zumino-Witten to Liouville \label{WZWL}}

Given the Chern-Simons form (\ref{aa2}) of the (2+1)-dimensional gravitational 
action, the arguments of section \ref{CSWZW} allow us to obtain a chiral 
$\mathrm{SL}(2,\mathbb{R})\times\mathrm{SL}(2,\mathbb{R})$ WZW action at the 
boundary.  In particular, for the BTZ black hole, the spacetime manifold has the 
topology $\mathbb{R}\times D^2$, and the WZW action describes the dynamics at the 
boundary $\mathbb{R}\times S^1$ at conformal infinity.  This appearance of a 
conformal field theory at the conformal boundary of an asymptotically anti-de 
Sitter space is perhaps the simplest example of the famous AdS/CFT correspondence 
of string theory \cite{Stromb}.

But as Coussaert, Henneaux, and van Driel first pointed out \cite{Coussaert}, we
have not yet exhausted the full set of boundary conditions implied by the asymptotic 
behavior (\ref{ab7}) of the connection.  The additional boundary conditions further
simplify the boundary theory, eventually reducing it to Liouville theory.

As a first step in this reduction, note that for the Chern-Simons connection 
(\ref{ab7}) describing the BTZ black hole, $A^{(+)}_v = 0 = A^{(-)}_u$.  This 
is exactly the type of boundary condition discussed in section \ref{CSWZW}, with 
$u\leftrightarrow z$ and $v\leftrightarrow {\bar z}$.  We thus obtain a sum of 
two chiral WZW models with opposite chiralities,
\beq
I = kI^+_{\mathit{WZW}}[g_1^{-1},{\bar A}_u=0] 
  - kI^-_{\mathit{WZW}}[g_2^{-1},{\bar A}_v=0] .
\label{cc1}
\eeq
But by the Polyakov-Wiegmann formula \cite{PW}, a pair of chiral WZW 
actions combines naturally to form a single nonchiral WZW action 
\beq
I = kI_{\mathit{WZW}}[g=g_1g_2^{-1}] .
\label{cc2}
\eeq

As a second step in the reduction, we can impose the additional conditions---which
also follow from (\ref{ab7})---that $A^{(+)}_u\sim\mathit{const.}\,T^-$ and 
$A^{(-)}_v\sim\mathit{const.}\,T^+$ at constant $\rho$.  These conditions 
translate into constancy of certain $\mathrm{SL}(2,\mathbb{R})$ components of 
two currents: after a $\rho$-dependent gauge transformation,  
\beq
\left(\partial_vg\,g^{-1}\right)^- = 1, \qquad
\left(g^{-1}\partial_ug \right)^+ = 1,
\label{cc3}
\eeq
where $X^\pm = \mathrm{Tr}(XT^\pm)$.  Constraints of this kind were first treated 
classically by Forgacs et al.\ \cite{Forgacs}, and, in a slightly different form, 
by Alekseev and Shatashvili \cite{AlekShat}; later the full quantum theory was 
analyzed by a number of authors \cite{Bershadsky,Papa,Balog,OR}.  The effect of 
the constraints is to eliminate two degrees of freedom, one directly and one because 
the conditions, treated as first class constraints, generate gauge transformations 
that can be factored out.  The classical derivation is straightforward: if we 
decompose the $\mathrm{SL}(2,\mathbb{R})$ group element $g$ as
\beq
g = \left(\begin{array}{cc}1&X\\0&1\end{array}\right)
    \left(\begin{array}{cc}e^{\varphi/2}&0\\0&e^{-\varphi/2}\end{array}\right)
    \left(\begin{array}{cc}1&0\\Y&1\end{array}\right) ,
\label{cc3a}
\eeq
the WZW action (\ref{cb8}) is simple to compute, yielding
\beq
kI^+ _{\mathrm{SL}(2,\mathbb{R})} = \frac{k}{4\pi}\int dudv\,\left[
     \frac{1}{2}\partial_u\varphi\partial_v\varphi 
     + 2e^{-\varphi}\partial_uX\partial_vY\right] ,
\label{cc3b}
\eeq
while the constraints (\ref{cc3}) become
\beq
\left(\partial_vg\,g^{-1}\right)^- = e^{-\varphi}\partial_vY = 1, \qquad
\left(g^{-1}\partial_ug \right)^+ = e^{-\varphi}\partial_uX = 1 .
\label{cc3c}
\eeq
Inserting these constraints into (\ref{cc3b}), we see that the action reduces to a 
Liouville action,
\beq
I = I_{\mathit{Liou}} = \frac{k}{8\pi}\int d^2x\sqrt{g}
  \left(\frac{1}{2}g^{ab}\partial_a\varphi\partial_b\varphi 
  +\frac{1}{2}\varphi R + \lambda e^\varphi\right) .
\label{cc4}
\eeq
In the original derivation of Coussaert et al., the metric in (\ref{cc4}) was
a flat metric on the cylinder at infinity, but a slight generalization of the
asymptotic conditions \cite{Rooman,Roomana} permits a general curved metric
and nontrivial holonomies.

Note that with fields normalized as in (\ref{cc4}), the coefficient in front of 
the Liouville action is $c/48\pi$, where $c$ is the classical central charge of 
the Virasoro algebra of the Liouville theory \cite{Seiberg}.  Thus, using
(\ref{aa3a}),
\beq
c = 48\pi\cdot\frac{k}{8\pi} = 6k = \frac{3\ell}{2G} ,
\label{cc5}
\eeq
in agreement with the Brown-Henneaux central charge (\ref{ba6}) obtained from 
the algebra of asymptotic symmetries.

Similar constructions exist for (2+1)-dimensional supergravity \cite{HMS,BanBaut}.
There are also hints that a discretized boundary Liouville theory can be obtained
from discretized (2+1)-dimensional gravity \cite{OLoughlin}.

The Liouville action (\ref{cc4}) depends on a background metric $g_{ab}$ at the
conformal boundary.  In both the Chern-Simons approach of \cite{Rooman,Roomana}
and the Fefferman-Graham construction described below in section \ref{FG}, a
conformal structure---that is, a conformal equivalence class of metrics---is 
prescribed as part of the boundary data.  One can make the dependence on the 
metric more explicit by starting with a fixed canonical metric and performing 
a quasiconformal deformation,
\beq
z\rightarrow f(z,{\bar z}) , \qquad {\bar z}\rightarrow {\bar f}(z,{\bar z}) ,
\label{cf2}
\eeq
which can change the conformal structure.  Such transformations have been considered 
in \cite{Guo,Guob}.  They are parametrized by a Beltrami differential $\mu$, such
that
\beq
\partial_{\bar z}f - \mu\partial_z f = 0 .
\label{cf3}
\eeq
The effect of the transformation (\ref{cf2}) is to change the Liouville action to 
a more general ``Beltrami-Liouville action,'' which includes Polyakov's light cone 
action for $f$ \cite{Polyakov} and which makes the dependence on the boundary 
conformal structure manifest.

One useful check on this derivation comes from looking at Ward identities.  It 
has been known for some time that the Virasoro Ward identities for Liouville
theory (in its nonlocal Polyakov form \cite{Polyakov,Polyakovb,Fujiwara}) can be 
derived from an $\mathrm{SL}(2,\mathbb{R})$ Chern-Simons theory \cite{Verlinde}, 
and thus, perhaps, from (2+1)-dimensional gravity \cite{Verlinde,Carlipi,Ashworth}.  
Ba{\~n}ados and Caro have recently shown how the full, nonchiral Liouville Ward 
identities can be obtained from asymptotically anti-de Sitter (2+1)-dimensional 
gravity \cite{BanCaro}.

\subsection{Asymptotic AdS and the Fefferman-Graham Construction \label{FG}}

The preceding section relied heavily on the Chern-Simons formulation of
(2+1)-dimensional gravity, and thus on special features peculiar to three 
dimensions.  It is naturally of interest to see whether a similar reduction 
to a boundary conformal field theory can be found in the standard metric 
formalism, perhaps allowing easier generalizations to more than 2+1
dimensions.

The key to such a reduction comes from a theorem of Fefferman and Graham 
\cite{FG,Kich}, who show that given a conformal metric at the boundary of an 
asymptotically anti-de Sitter spacetime, there exists a (formal) asymptotic 
expansion of the metric that solves the vacuum Einstein field equations.  In 
particular, in three spacetime dimensions, one can find coordinates such 
that 
\beq
ds^2 = \frac{\ell^2}{r^2}dr^2 + \frac{r^2}{\ell^2}g_{ij}(r,x)dx^idx^j 
 \quad \hbox{with}\quad g_{ij}(r,x) = \overset{(0)}g_{ij}(x) 
 + \frac{\ell^2}{r^2}\overset{(2)}g_{ij}(x) + \frac{\ell^4}{r^4}\overset{(2)}g_{ij}(x) ,
\label{cd1}
\eeq
where indices $i,j$ run from $1$ to $2$.\footnote{In general, one obtains an
infinite series for $g_{ij}$, which includes logarithmic terms as well; but in
three dimensions, the series terminates \cite{SkenSolo}, as is evident in the
general solution (\ref{ab5}).}  The Einstein field equations then become 
\cite{HennSken,Rooman,SkenSolo,Bautier,Bautierb}
\begin{align}
&\overset{(2)}g\,{}^i{}_i = -\frac{\ell^2}{2}\overset{(0)}R \nonumber\\
&\overset{(0)}\nabla_i\overset{(2)}g_{jk} -
 \overset{(0)}\nabla_j\overset{(2)}g_{ik} = 0 ,
\label{cd2}
\end{align}
where indices are raised and lowered and covariant derivatives defined in terms 
of the conformal boundary metric $\overset{(0)}g_{ij}$.  Note that these equations
can be obtained directly from the Hamiltonian and momentum constraints of
(2+1)-dimensional gravity, without using the full ``bulk'' field equations.

As a first step, we can now reproduce the conformal anomaly described at the
end of section \ref{AS} \cite{HennSken,Bautierb}.  The coordinate transformations 
that preserve the form (\ref{cd1}) of the metric are \cite{Bautierb,Imbibo}
\begin{align}
\delta r &= -r\delta\sigma(x) \nonumber\\
\delta x^i &= \int_0^r \frac{\ell^2}{r^{\prime 3}}g^{ij}(x,r')\partial_j\sigma dr',
\label{cd3}
\end{align}
and it is straightforward to check that the corresponding change in the metric
is
\begin{align}
\delta\overset{(0)}g_{ij} &= -2\delta\sigma\overset{(0)}g_{ij} \nonumber\\
\delta\overset{(2)}g_{ij} 
  &= -\ell^2\overset{(0)}\nabla_i\overset{(0)}\nabla_j\delta\sigma .
\label{cd4}
\end{align}
The boundary term in the Einstein-Hilbert action (\ref{ba12}) is chosen in
such a way that the boundary variation vanishes when the metric is held
fixed at the boundary.  We therefore know from general principles---and also,
of course, from explicit computation---that
\beq
\delta I_{\mathit{EH}} = -\int_{\partial U}d^2x\,\pi_{ij}\delta\gamma^{ij} ,
\label{cd4a}
\eeq
where in three dimensions the canonical momentum $\pi_{ij}$ is \cite{Carlipb}
\beq
\pi_{ij} = \frac{1}{16\pi G}\sqrt{\gamma}\left( K_{ij} - \gamma_{ij}K\right) .
\label{cd4b}
\eeq
The counterterm (\ref{ba13}) also varies in an easily calculable fashion.  Then
using (\ref{cd2}) and (\ref{cd4}), it is straightforward to show that
\beq
\delta I 
 = \frac{1}{8\pi G\ell}\int_{\partial U}d^2x \sqrt{-\overset{(0)}g}\,\delta\sigma 
 \overset{(2)}g\,{}^i{}_i = -\frac{\ell}{16\pi G}\int_{\partial U}d^2x
 \sqrt{-\overset{(0)}g}\,\delta\sigma \overset{(0)}R ,
\label{cd5}
\eeq
which can be recognized as the standard expression for the conformal anomaly of 
a two-dimensional field theory with central charge $c=3\ell/2G$ \cite{CFT}.
More generally, the variation (\ref{cd4a}) can be used to extract the 
``holographic stress tensor'' (\ref{cd7}) at the boundary \cite{Haro,Papadimitriou,%
BanSchTh,Papadimitrioub}.

We can now obtain Liouville theory in several ways.  The most direct 
\cite{Bautier,Bautierb} is to simply postulate that $\overset{(2)}g_{ij}$,
which has one independent degree of freedom, can be written in terms of
a scalar field $\varphi$ such that $e^\varphi$ has conformal weight $-1$ (that is,
under a Weyl transformation $\delta\varphi = -\sigma$).  It is not hard to
show that the combination that has the transformation property (\ref{cd4})
is
\beq
\overset{(2)}g_{ij} = \ell^2\left[ -\overset{(0)}\nabla_i\overset{(0)}\nabla_j\varphi
 + \overset{(0)}\nabla_i\varphi\overset{(0)}\nabla_j\varphi + \overset{(0)}g_{ij}
 \left(\lambda e^{2\varphi} - \frac{1}{2}\overset{(0)}g{}^{kl}
 \overset{(0)}\nabla_k\varphi\overset{(0)}\nabla_l\varphi\right)\right] ,
\label{cd6}
\eeq
where $\lambda$ is an arbitrary constant.  This expression is closely related to 
the stress-energy tensor of Liouville theory.  Indeed, the Einstein equations
(\ref{cd2}) imply that
\beq
\overset{(0)}\nabla_i T^i{}_j = 0 \qquad \hbox{with} \quad
T^i{}_j = \frac{1}{8\pi G\ell}\left(
 \overset{(2)}g{}^i{}_j + \frac{\ell^2}{2}\delta^i_j\overset{(0)}R\right) ,
\label{cd7}
\eeq
and with the identification (\ref{cd6}), $T^i{}_j$ is precisely the stress-energy
tensor obtained from the Liouville action (\ref{cc4}).

A roughly equivalent procedure \cite{SkenSolo} is to use a Liouville field as
an auxiliary field in order to directly integrate the Einstein equations (\ref{cd2})
or (\ref{cd7}).  Since these are differential equations, $\overset{(2)}g_{ij}$ 
will be a nonlocal function of the prescribed data $\overset{(0)}g_{ij}$.  One can, 
however, restore locality by
introducing an auxiliary field $\varphi$.  The expression (\ref{cd6}) can then be
viewed as simply giving an integral of (\ref{cd7}); the nonlocality is now
hidden in the fact that $\varphi$ must itself obey the Liouville equation of motion, 
and thus depends nonlocally on $\overset{(0)}g_{ij}$.  Alternatively, one can
``integrate the anomaly''; that is, one can look directly for an action depending 
on the boundary metric whose conformal variation is given by (\ref{cd5}) 
\cite{SkenSolo,Schwimmer,BanChan}.  The result has a unique nonlocal piece, the 
Polyakov action \cite{Polyakov,Polyakovb}
\beq
I_{\mathit{Pol}} = \int d^2x \sqrt{\gamma}R\,\Box^{-1}R ,
\label{cd8}
\eeq
which is essentially equivalent to the Liouville action \cite{Fujiwara}.

Such nonlocality should not be surprising, in view of our picture of the dynamical
degrees of freedom as ``would-be gauge'' excitations.  The great strength of the
Chern-Simons formalism is that the gauge transformations are local.  In the metric
formalism, on the other hand, gauge transformations---diffeomorphisms---are not
local: after all, a diffeomorphism moves points.  The surprise is not that the
dynamical description of boundary diffeomorphisms is nonlocal, but rather that it 
is ``local enough'' to allow an easy description.

We have not yet directly related our Liouville field to the asymptotic
diffeomorphisms of the black hole metric.  This can be done \cite{Roomanb,%
Carlipj,Krasnovb}, essentially by considering the finite version of the
transformation (\ref{cd3}).  Let $\rho=\ln r$.  Under a diffeomorphism%
\footnote{The exact version of this transformation, to valid all orders in 
$e^{-2\rho}$, may be found in \cite{Roomanb,Krasnovb}.}
\begin{align}
\rho &\rightarrow \rho + \frac{1}{2}\varphi(x) + e^{-2\rho}\overset{(2)}f(x) 
  + \dots \nonumber\\
x^i &\rightarrow x^i + e^{-2\rho}\overset{(2)}h{}^i(x) + \dots ,
\label{cd9}
\end{align}
the demand that the metric remain in the form (\ref{cd1}) leads to the
relations \cite{Skenx}
\beq
\overset{(2)}h_i = -\frac{\ell^2}{4}e^{-\varphi}\partial_i\varphi , \qquad
\overset{(2)}f = -\frac{\ell^2}{16}e^{-\varphi}\,\overset{(0)}g{}^{ij}
  \partial_i\varphi\partial_j\varphi .
\label{cd10}
\eeq
It is then easy to show that the spatial metric transforms to (\ref{cd6})
with $\lambda=0$.  But the field $\varphi$ is no longer an auxiliary field
introduced to integrate the field equations; rather, it is an explicit
parametrization of the asymptotic diffeomorphisms (\ref{cd9}).  

One can take this argument a step further, and compute an action for $\varphi$.
In the original Einstein-Hilbert action (\ref{ba12}) with the counterterm
(\ref{ba13}), we are instructed to integrate out to a constant value 
$\rho= \bar\rho$, sum the terms in the action, and only then take the
limit $\bar\rho\rightarrow\infty$.  With the coordinate transformation
(\ref{cd9}), though, we should place the boundary at a location at which
the \emph{new} radial coordinate is constant; that is, in the original 
coordinate system,
\beq
\rho = \bar\rho + \frac{1}{2}\varphi + e^{-2\bar\rho}\overset{(2)}f 
  + \dots  .
\label{cd11}
\eeq
A straightforward calculation then shows that the Einstein-Hilbert action
reduces to the Liouville action (\ref{cc4}) at the boundary \cite{Manvelyan,Carlipj}.
Moreover, by comparing the BTZ metric (\ref{ab5}) with the expansion (\ref{cd6})
of the metric, we can confirm that the constants $L^\pm$ of (\ref{ab6}) are
indeed the classical values of the Virasoro generators $L$ and $\bar L$ of
the Liouville theory.  We thus confirm the picture of the boundary degrees of 
freedom as ``would-be diffeomorphisms'' that become dynamical at the conformal 
boundary.\footnote{See also \cite{SkenSolo} for the case of a static boundary
metric; that paper also discusses some important subtleties involving the lower 
range of the integration over $\rho$ in the bulk action.}

It is worth noting that the metric derivations of Liouville theory typically
lead to an action in which the constant $\lambda$ either vanishes or is put in 
by hand.  The Chern-Simons derivation, on the other hand, naturally leads to a
nonzero value of $\lambda$.  The difference arises from the difference 
in boundary conditions.  Rather than fixing the metric at the boundary, the 
Chern-Simons derivation fixes components of the connections $A^{(\pm)}$.  This 
change has two consequences \cite{BanMend}: the coefficient of the extrinsic 
curvature term in the action (\ref{ba12}) is altered, and the extrinsic curvature 
is itself replaced by a first-order form,
\beq
\int d^2x\sqrt{\gamma}{\tilde K} = \int\omega_a\wedge e^a .
\label{cd11a}
\eeq
One can always find a local Lorentz frame in which $\tilde K = K$.  But
$\tilde K$ is not invariant under local Lorentz transformations, and new 
degrees of freedom---``would-be local Lorentz transformations''---appear.
The Liouville potential term in the Chern-Simons derivation can be traced
directly to these extra degrees of freedom \cite{Carlipj}.

This difference may not be as large as it first appears, however.  As noted 
in a different context in \cite{BanChan}, the vanishing of $\lambda$ in the 
metric formalism comes from an exact cancellation between the divergence 
of the Einstein-Hilbert action and the boundary counterterm (\ref{ba13}).  
If this term is interpreted as in conventional renormalization, we might 
instead expect a finite remainder, which, it has been argued, can give 
precisely the missing potential term in the Liouville action.   

\subsection{Euclidean Gravity and Liouville Theory}

If we allow ourselves to analytically continue from Lorentzian to Riemannian
metrics, there is another more manifestly geometrical way of obtaining 
Liouville theory from the asymptotic behavior of three-dimensional gravity.  
This Euclidean theory has been developed in depth by Krasnov \cite{Krasnovb,%
Krasnova,Krasnovc,Krasnovd,Krasnove,Krasnovf,Krasnovg}; here I will merely
summarize some key features.

Any solution of three-dimensional (Euclidean) gravity with a negative cosmological 
constant is a constant negative curvature space, and can be expressed as a quotient
of hyperbolic three-space $\mathbb{H}^3$ by discrete groups of isometries,
$M=\mathbb{H}^3/\Gamma$.  Equivalently, one can view $M$ as a piece of 
$\mathbb{H}^3$ with boundaries ``glued'' together.  The Euclidean continuation 
of the BTZ black hole, for example, is a quotient of $\mathbb{H}^3$ by a group 
generated by a single element \cite{CarTeit}; in the upper half-space model 
of $\mathbb{H}^3$, 
\beq
ds^2 = \frac{\ell^2}{z^2}\left( dx^2 + dy^2 + dz^2\right), \qquad z>0,
\label{ce1}
\eeq
the identification is just a dilatation combined with a rotation around the
$z$ axis, with parameters determined by the mass and angular momentum of
the black hole.  By using more complicated isometries, this picture can be 
extended to more elaborate configurations; for example, there are Euclidean 
versions of multiple-horizon black holes, and of black holes combined with point 
particles.

One can now compute the Einstein-Hilbert action for such configurations.
As in the preceding section, the result is divergent, with both an area and a 
logarithmic divergence that must be regulated.  The delicate issue is how to 
choose a boundary analogous to ``constant $\rho$'' upon which to perform 
the regularization, in such a way that the identifications $\Gamma$ act nicely.  
Specifically, one would like a boundary whose metric approaches the standard 
upper half-space metric of $\mathbb{H}^2$ as the regulator is removed.  The 
answer is given by classical geometry: one should choose a surface 
\cite{Krasnova}
\beq
z(x,y) = \epsilon e^{-\varphi(x,y)}
\label{ce2}
\eeq
where $\varphi(x,y)$ is a solution of the Liouville equation.  Given such a choice
of boundary, it can be shown that the regulated Einstein-Hilbert action is
precisely the Liouville action (\ref{cc4}), with additional boundary terms
\cite{TakZog} related to the action of the identifications $\Gamma$ on the 
boundary.

It is clear that this result should be related to the ``would-be diffeomorphism''
description \cite{Carlipj} discussed at the end of the preceding section.  In
particular, (\ref{ce2}) is probably at least roughly the Euclidean equivalent  
of the asymptotic diffeomorphism (\ref{cd9}), although details have not yet been 
worked out.  These results may also offer further insight into the missing
``potential term'' discussed at the end of that section.  The Liouville field in
(\ref{ce2}) comes from uniformization of the boundary surface, that is, from
a Weyl transformation to the canonical metric of constant curvature at the
boundary.  For a single BTZ black hole, the Euclidean boundary is a torus,
and the canonical metric is flat; as a result, the classical Liouville equation 
has $\lambda=0$.  For more complicated topologies---multiple-horizon solutions,
for example---the Euclidean boundary is a surface of genus $g>1$, the canonical
metric is one of constant negative curvature, and the classical Liouville equation 
has a nonvanishing $\lambda$.  This suggests that an extension of \cite{Carlipj}
to multiple-horizon black holes might automatically introduce a nontrivial 
value of $\lambda$.

The boundary Liouville action described here is relatively new, but it has 
already lead to a number of interesting applications.  In particular, the 
formalism has been used to compute the thermodynamics of multiple-horizon black 
holes \cite{Krasnovg} (a Cardy formula interpretation of these results would be 
very interesting) and to analyze the quantum production of black holes by point 
particle collisions \cite{Krasnovf}.  The formalism also gives a Liouville theory 
expression for the probability of point particle emission by a BTZ black hole 
\cite{Krasnove} that appears to be at least qualitatively correct (see also 
\cite{Solo}).

\subsection{Projective Structures}

The asymptotic dynamics of (2+1)-dimensional anti-de Sitter gravity certainly 
appears to contain Liouville theory, but there are some hints that additional 
structure may be necessary as well.  The general form (\ref{ab5}) of the
asymptotic metric incorporates both a complex structure\footnote{For a single 
Lorentzian black hole, the complex structure is almost, but not quite, trivial: 
it is determined by the relative scale of $t$ and $\phi$ in the coordinates $u$, 
$v$ of (\ref{ab5}).  For a multi-black hole solution, the choice is generally 
much more complicated.  Note that for an orientable two-manifold, a complex
structure and a conformal structure are equivalent; see \cite{Nelson} for a
good physicists' introduction.} and a pair of functions $L^\pm$.  As noted in 
\cite{Ezawab,Nakatsu}, the latter combine to form a holomorphic quadratic 
differential, essentially a transverse traceless rank two tensor.  Such data, 
in turn, determine a projective structure \cite{Gunning} at infinity, that is, 
a collection of complex coordinate patches whose transition functions are 
fractional linear transformations.  The space of projective structures is 
closely related to the cotangent bundle of the space of conformal structures,
or Teichm{\"u}ller space; a short introduction may be found in Appendix C of 
\cite{Krasnovb}.

The derivations of Liouville theory described above have treated the $L^\pm$ as 
functions of the Liouville field $\varphi$, thus reducing the dynamics to that 
of a single field with a single background metric.  But this may not be sufficiently 
general.  Krasnov has argued in \cite{Krasnovb} that the full space of projective 
structures is the correct moduli space of classical multi-black-hole solutions, 
at least in the Euclidean setting, and that the inclusion of projective structures 
in the path integral leads to good holomorphic factorization properties for the 
Euclidean partition function.  The space of projective structures is twice as 
large as the space of conformal structures, and one may argue that this is the 
``right'' size.  A single $\mathrm{SL}(2,\mathbb{R})$ Chern-Simons theory gives 
a partition function $Z[\mu,\bar\mu]$ that depends on a Beltrami differential 
(\ref{cf3}) and can be interpreted as a ``quantization of Teichm{\"u}ller 
space'' \cite{Verlinde}; it is not unreasonable to expect the partition function 
of an $\mathrm{SL}(2,\mathbb{R})\times\mathrm{SL}(2,\mathbb{R})$  Chern-Simons 
theory to be, in some sense, twice as big.  Indeed, the derivation of Ward 
identities in \cite{BanCaro} seems to require such a treatment of the $L^\pm$.

In the Fefferman-Graham approach of section \ref{FG}, it is possible to see where
this extra structure might be hidden.  There, the metric coefficients $L^\pm$ were 
determined as functions of the Liouville field by integrating the asymptotic
Einstein equation (\ref{cd7}).  But the solution (\ref{cd6}) is not unique: as
noted in \cite{SkenSolo,Krasnovx}, one can add to $T^i{}_j$ any solution 
${\tilde T}^i{}_j$ of
\beq
\nabla_i{\tilde T}^i{}_j = 0, \qquad {\tilde T}^i{}_i=0 ,
\label{cf1}
\eeq
that is, any holomorphic quadratic differential.  It has been pointed out in 
\cite{BanChan} that a shift arises naturally when one considers asymptotically 
nontrivial diffeomorphisms: because of the anomalous transformation properties 
(\ref{ba10}), a holomorphic diffeomorphism can shift $L^+(u)$ by an arbitrary 
function of $u$ and $L^-(v)$ by an arbitrary function of $v$.   

Despite these intriguing results, though, the role of projective structures in 
(2+1)-dimensional asymptotically anti-de Sitter gravity remains fairly mysterious. 
To the best of my knowledge, no one has yet systematically studied their role, and
there has been no work on quantization in this context.

\section{Counting States \label{count}}
\setcounter{footnote}{0}

We have now answered the first question raised in the Introduction: although
(2+1)-dimensional gravity on a spatially compact manifold is ``topological,'' 
on a noncompact manifold the theory acquires a new set of dynamical degrees 
of freedom, described by an $\mathrm{SL}(2,\mathbb{R})\times\mathrm{SL}(2,\mathbb{R})$ 
WZW action or a Liouville action.  The next question is whether we can count
these degrees of freedom to obtain a microscopic explanation of the entropy of
the BTZ black hole.

This is, unfortunately, a far more difficult question than one might naively expect.  
WZW models are well understood, but only for compact gauge groups; noncompact groups
such as $\mathrm{SL}(2,\mathbb{R})$ are much harder to handle.  Similarly,
one sector of Liouville theory, the ``normalizable'' sector, is reasonably well
understood \cite{Seiberg}, but the degrees of freedom relevant to the BTZ black
hole come mainly from the much more poorly understood ``nonnormalizable'' sector.
Nevertheless, some progress has been made in counting states, and there is hope
that we may soon understand the problem better.

\subsection{WZW Approaches: General Considerations \label{WGC}}

An obvious first step towards counting BTZ black hole states is to try to understand 
the states of the $\mathrm{SL}(2,\mathbb{R})\times\mathrm{SL}(2,\mathbb{R})$ WZW
model induced at the conformal boundary.  WZW models have been studied extensively
(see, for example, \cite{CFT}), and in many cases their states and operators are
well understood.  Unfortunately, though, WZW models with noncompact groups are much
more poorly understood; although there has been some progress---see, for example,
\cite{Balogb,Dixon,BarNatan,Wittend,Hayashi,Chaudhuri,Bars,Nichols,Bertoldi,MalOoga,%
MalOogb,MalOogc}---we remain largely ignorant.  

If we wish to explain black hole entropy by counting WZW states, we are
immediately confronted with a problem.  In general, a WZW model for group $G$ 
is characterized by an affine current algebra (\ref{da1}), with Virasoro operators
given by (\ref{da2}).  The central charge may be computed directly by evaluating the commutators of the $L_m$; for $\mathrm{SL}(2,\mathbb{R})$, one obtains
\beq 
c = \frac{3k}{k-2} .
\label{da3}
\eeq
In particular, in the semiclassical ($k\gg1$) limit, $c\approx3$.  One can
understand this limit by rescaling the currents in (\ref{da1}) to ${\tilde J} =%
J/\sqrt{k}$.  For large $k$, the term involving the structure constant is suppressed, 
and one obtains three decoupled $\mathrm{U}(1)$ current algebras, each having 
$\tilde k = 1$.  But the resulting central charge is then \emph{much} smaller than 
the Brown-Henneaux value (\ref{ba6}), and the Cardy formula would seem to give an 
entropy that is drastically too small.

But the Cardy formula also suggests a possible escape \cite{Carlipg}.  The central 
charge appearing in the expression for the density of states is actually the 
effective central charge (\ref{bb1}).  If we can find a model for which the lowest 
eigenvalue of $L_0$ is
\beq
\Delta_0 = -\frac{k}{4},
\label{da4}
\eeq
then from (\ref{bb1}) and (\ref{aa3a}), we would obtain $c_{\hbox{\scriptsize\it eff}}
\approx 6k = 3\ell/2G$, matching the central charge of the asymptotic algebra. 
Of course, a negative value of $\Delta_0$ would ordinarily imply a nonunitary
theory, and while this is not necessarily fatal \cite{Cohn,Gannon}, it may seem
unnatural.  On the other hand, $\mathrm{SL}(2,\mathbb{R})$ WZW models already have
problems related to unitarity \cite{Balogb,Dixon}, so this might not be so terrible.

In fact, the choice (\ref{da4}) has some justification, both from gravity and from
conformal field theory.  On the gravity side, if one normalizes the Virasoro
algebra so that the zero-mass BTZ black hole has $L_0=0$, then there exists a large 
class of point particle states for which the classical value of $L_0$ is negative
\cite{BrownHenn}, with values ranging from $0$ down to precisely $-k/4$.  The
lowest eigenvalue (\ref{da4}) corresponds to empty anti-de Sitter space.  The
relevance of such point particle states receives some support from their apparent
relation to the quasinormal modes of the BTZ black hole \cite{BirmCarChen}, and
we shall see below in section \ref{LT} that closely related states may be quite
important in understanding quantum Liouville theory.  Classically, point particles
with $L_0<0$ can collide to form black holes \cite{Matschull}, and Krasnov has
suggested that perhaps the entropy of the BTZ black hole can be viewed as counting
possible point particle constituents \cite{Krasnovc}.

It is also worth noting that in supergravity, the massless black hole is the
vacuum state of the Ramond sector of the theory, while anti-de Sitter space
is the vacuum state of the Neveu-Schwarz sector \cite{CousHenn}.  The $L_0$
eigenvalues of these two vacua differ by $c/24=k/4$, in agreement with the
comments above.  Unfortunately, this argument does not fix an absolute value
of $\Delta_0$, since the superconformal algebra may be easily deformed in a
way that shifts both $\Delta_0$ and $c$ \cite{Carlipg}.

The value (\ref{da4}) also has a (somewhat weaker) rationale from the conformal 
field theory side.  The Virasoro generator $L_0$ of equation (\ref{da1}) 
involves an important zero-mode term:
\beq
L_0 = \frac{1}{k-2}\left( (J_0^0)^2 - (J_0^1)^2 - (J_0^2)^2 \right) + N ,
\label{da5}
\eeq
where
\beq
N = \frac{2}{k-2}\sum_{m=1}^\infty\left( J^0_{-m}J^0_{m} - J^1_{-m}J^1_{m} 
  - J^2_{-m}J^2_{m}\right)
\label{da5a}
\eeq
is essentially the sum of three number operators.
From the representation theory\footnote{For a description of the representations
of $\mathrm{SL}(2,\mathbb{R})$, see, for example, \cite{Inomata,Lang,Troost};
a brief summary appears below in section \ref{QWZW}.} of $\mathrm{SL}(2,\mathbb{R})$, 
we learn that for the principal discrete series,
\beq
L_0 = - \frac{j(j+1)}{k-2} + N ,
\label{da6}
\eeq
where $j$ is a negative integer or half-integer.  In particular, for $j=-k/2$,
\beq
L_0 = -\frac{k}{4} + N .
\label{da7}
\eeq

Now, in an $\hbox{SU}(2)$ Chern-Simons theory, $j=k/2$ is the highest admissible 
value (the highest integrable representation), and Hwang has argued that $j=-k/2$ 
may play an equivalent role for $\mathrm{SL}(2,\mathbb{R})$ \cite{Hwang,Hwang2,Evans}. Similarly, in the Euclidean partition function approach of \cite{Carlipk}, 
$|j|=|k/2|$ is the maximal value appearing in the partition function.  While 
certainly none of this is conclusive in the absence of a carefully worked out 
quantization of the $\mathrm{SL}(2,\mathbb{R})$ WZW model, these results are at 
least suggestive.

An alternative approach to the ``low central charge problem,'' proposed independently 
and in slightly different forms by Ba{\~n}ados \cite{Banadosc,Banadosx} and Fjelstad 
and Hwang \cite{FjelHwang}, is to allow some or all of the currents (\ref{da1}) to 
have fractional modes,
\beq
J^a = \sum_n \sum_{p=0}^{N-1} J_{n + \frac{p}{N}}^a 
  \exp\left\{i\left(n + \frac{p}{N}\right)\phi\right\} .
\label{da8}
\eeq
In general, this changes the physics; currents with fractional modes correspond
to metrics with certain conical singularities.  This may be reasonable from the
point of view of Liouville theory; Krasnov \cite{Krasnovc} and Chen \cite{Chen}
have both argued for an important role for point particles in understanding
the states of the BTZ black hole.

Once one has decided to admit such conical singularities, though, it is not 
obvious how---or whether---one should pick a particular value of $N$, that is, 
why one should still demand periodicity under transformations
\beq
\phi\rightarrow\phi+2\pi N . 
\label{da9}
\eeq
Ba{\~n}ados has argued that such a periodicity for the Virasoro generators, with 
$N=c$, can come from demanding unitarity of the Virasoro algebra \cite{Banadosc,%
Banadosx}.  Given a Virasoro algebra with central charge $c$, one may check that
a transformation
\beq
L_n \rightarrow {\tilde L}_n = NL_{n/N}
\label{da9a}
\eeq
gives a new Virasoro algebra with central charge ${\tilde c} = c/N$.  Requiring
that $N=c$ ensures that this new algebra is unitary; apart from a few exceptional
values of $N$ and $c$, this will not be the case if $N>c$ \cite{CFT}.   
Alternatively, looking at the full $\mathrm{SL}(2,\mathbb{R})$ current 
algebra, Fjelstad and Hwang have argued that a similar periodicity, with $N=k/2$, 
can come from the condition that the current zero-mode $J^0_0$ have integer 
eigenvalues \cite{FjelHwang}.  Fractional periodicity of the currents $J^a$ 
implies a fractional ``spectrally flowed sector'' similar to those described below 
in section \ref{QWZW}; the Fjelstad-Hwang condition follows from (\ref{db7}).  In
either case, plausible combinatorial arguments then lead to the correct BTZ
black hole entropy.

Yet another speculative approach to nonintegral moding has appeared
in \cite{Carlipg}.  Instead of only allowing states built from the vacuum
by currents $J^a_{-n}$ as in (\ref{db5}), one might also consider states 
formed by acting on a suitable vacuum by the group-valued WZW field $g$.  The 
conformal weight of $g$ is not integral: for a discrete representation of spin $j$,
\beq
\Delta_j(g) = \frac{j(j-1)}{k-2} ,
\label{db11}
\eeq
offering a new source for fractional conformal weights.  Indeed, if we assume that 
all spins appear and that representations with spin $j$ occur with a multiplicity 
$2j+1$, as suggested by the Plancherel measure \cite{Hwang2}, the combinatorial 
formula (\ref{bb7}) yields a density of states \cite{Carlipg}
\beq
\ln\rho(\Delta) \sim 2\pi\sqrt{\frac{k\Delta}{3}} .
\label{db14}
\eeq
This is the right order of magnitude, but differs from the correct answer
(\ref{bb8}) by a factor of $1/\sqrt3$.  The missing factor may reflect the 
``bimodular'' properties of the WZW model \cite{Chau}---$g$ transforms on one 
side under the standard $\mathrm{SL}(2,\mathbb{R})$ Lie algebra and on the 
other side under a quantum group, and one may speculate that the ``extra'' 
quantum group transformation leads to a further degeneracy in the number of 
states within a given representation of $\mathrm{SL}(2,\mathbb{R})$.

\subsection{Counting WZW States}

Despite the absence of a complete quantization of the $\mathrm{SL}(2,\mathbb{R})$ 
WZW model, a number of attempts have been made to count WZW states.  The early
efforts circumvented the difficulties arising from the noncompactness of 
$\mathrm{SL}(2,\mathbb{R})$ by, implicitly or explicitly, looking at analytic 
continuations to models with more compact-group-like behavior.  

The first attempt at such a counting of states, Ref.\ \cite{Carlipl}, looked at 
states that are exactly diffeomorphism-invariant, and in particular obey the 
condition $L_0|\mathit{phys}\rangle=0$.  This equality is achieved by balancing 
the negative contribution of the zero-modes in (\ref{da5}) with positive non-zero-%
mode ``oscillator'' contributions, with the implicit assumption that each component 
of $:g_{ab}J^a_mJ^b_{-m}:$ in (\ref{da2}) makes a \emph{positive} contribution, 
despite the indefiniteness of the metric.  Unlike most later papers, \cite{Carlipl} 
looked at the WZW model induced at the horizon rather than at infinity.  This 
changes the natural choice of boundary data, and one must perform a functional 
Legendre transformation to recover the usual partition function and density of 
states.  Details of this transformation may be found in Appendix B of \cite{Carlipg},
while \cite{BanGomb} contains a more careful derivation of the boundary conditions.  
Given these assumptions, though, one obtains precisely the correct Bekenstein-Hawking 
entropy (\ref{ab9}).  A similar computation gives the correct entropy for 
(2+1)-dimensional de Sitter space \cite{Stromc}.  In retrospect, the success of 
this approach is a bit surprising, since a crucial step depends on a cancellation 
between quantum corrections in the two $\mathrm{SL}(2,\mathbb{R})$ factors of the 
gauge group.

A second paper, Ref.\ \cite{Carlipk}, made an explicit analytic continuation 
from $\mathrm{SL}(2,\mathbb{R})\times\mathrm{SL}(2,\mathbb{R})$ to a
``Euclidean'' $\mathrm{SL}(2,\mathbb{C})$ Chern-Simons theory, whose
partition function is fairly well understood \cite{Wittend,Hayashi}:
\beq
Z_{\hbox{\scriptsize SL}(2,\mathbb{C})}[{\bar A}^+, {\bar A}^-]
  = \left| Z_{\hbox{\scriptsize SU}(2)}[\bar A] \right|^2 , 
\label{db1}
\eeq
where the $\mathrm{SU}(2)$ partition function $Z_{\hbox{\scriptsize SU}(2)}[\bar A]$
on a torus of modulus $\tau$ coupled to a background field ${\bar A}_z$ can be 
written explicitly in terms of Weyl-Kac characters for affine $\mathrm{SU}(2)$.  
The density of states can be extracted from the partition function (\ref{db1}) 
by a contour integral, since 
\beq
Z_{\hbox{\scriptsize SL}(2,\mathbb{C})}(\tau)[{\bar A}^+, {\bar A}^-]
  = \mathrm{Tr}\left\{ e^{2\pi i\tau L_0}e^{-2\pi i\bar\tau \bar L_0}\right\}
  = \sum \rho(\Delta,\bar\Delta)q_1^{\Delta-\bar\Delta}q_2^{\Delta+\bar\Delta} ,
\label{db2}
\eeq
where $q_1=e^{2\pi i\tau_1}$, $q_2=e^{-2\pi\tau_2}$, and $\rho(\Delta,\bar\Delta)$
is the number of states for which the Virasoro generators $L_0$ and $\bar L_0$ 
have eigenvalues $\Delta$ and $\bar\Delta$.  The zero-modes again play a crucial 
role: the partition function $Z_{\hbox{\scriptsize SU}(2)}[\bar A]$ is dominated 
by a zero-mode contribution coming from the coupling to the fixed boundary data 
${\bar A}_z$ in (\ref{cb7}) and (\ref{cb8}).  This prefactor gives the leading 
contribution to the density of states $\rho(\Delta,\bar\Delta)$, which again 
reproduces the Bekenstein-Hawking entropy (\ref{ab9}).  A similar argument has
appeared in \cite{BanOrt}, and an extension to (2+1)-dimensional de Sitter space 
has also been found in \cite{BanOrtc}.

Unfortunately, like most Euclidean path integral methods, the computation of
\cite{Carlipk} gives us the correct entropy without actually telling us much about 
the microscopic states responsible for that entropy.  There are some hints---for
example, the continuation requires that we flip the sign of the coupling constant $k$, 
a transformation that also shows up in some analyses of the $\mathrm{SL}(2,\mathbb{R})$
WZW partition function \cite{Henn}.  To go further, though, we can no longer avoid 
the task of quantizing the $\mathrm{SL}(2,\mathbb{R})$ WZW model.  

\subsection{Toward a Quantum $\mathbf{SL}(2,\mathbb{R})$ WZW Model \label{QWZW}}

I believe it is fair to say that no one yet fully understands how to quantize
the $\mathrm{SL}(2,\mathbb{R})$ WZW model.  There has been a good deal of work on 
this problem in the past several years, though, particularly in the context of string 
theory and the AdS/CFT correspondence.  My treatment here will be sketchy, but I
will try to summarize the major progress and open questions.

Quantization of the algebra (\ref{da1}) requires two steps: we must choose a ``vacuum''
representation for the algebra of the zero-modes $J^a_0$, and must then build up a
representation of the rest of the $J^a_n$, which should be ghost-free (that is,
states should have positive norms) and should allow the construction of a modular 
invariant partition function.  The first step is classical: the representation theory
of $\mathrm{SL}(2,\mathbb{R})$ was analyzed by Bargmann in 1947 \cite{Bargmann}, and
is discussed in, for example, \cite{Inomata,Lang,Troost}.  As in the better-known
case of $\mathrm{SU}(2)$, one can find a basis of simultaneous eigenfunctions of
$J^0_0$ and $c_2 = \eta_{ab}J^a_0J^b_0$, with\footnote{Note that conventions vary; 
the equations below depend on the choice of signature of the metric.}
\beq
J^0_0 |j,m\rangle = m|j,m\rangle ,\qquad c_2|j,m\rangle = -j(j+1)|j,m\rangle .
\label{db3}
\eeq
Unlike the case of $\mathrm{SU}(2)$, the unitary representations are all infinite
dimensional, and come in a number of classes:
\begin{enumerate}
\item the discrete representations $\mathcal{D}^\pm_j$, $j<0$, $-2j \in\mathbb{Z}$;
\item the principal continuous representations $\mathcal{C}^\epsilon_j$, 
$\epsilon = 0,1/2$, $j = -\frac{1}{2} + is$ with $s$ real;  
\item the complementary representations $\mathcal{E}_j$, $-1<j<0$;
\item the identity representation.
\end{enumerate}
Additional possibilities appear if one considers the universal covering of
$\mathrm{SL}(2,\mathbb{R})$; in particular, various quantities need no longer 
be integers.  One can also obtain a different set of ``hyperbolic'' representations
by simultaneously diagonalizing $J^2_0$ and $c_2$ \cite{Lindblad}; in contrast to 
the case of $\mathrm{SU}(2)$, the signature of the Cartan-Killing metric now
distinguishes $J^0$ from $J^1$ and $J^2$.  I will mention these representations
briefly at the end of this section.

It is not obvious \emph{a priori} which representations should be relevant to the
$\mathrm{SL}(2,\mathbb{R})$ WZW model.  Only the discrete and continuous representations 
occur in the Peter-Weyl decomposition of square integrable functions on the group 
manifold, and it has been argued that these are therefore the only relevant 
representations for string theory \cite{Evans,MalOoga}.  If we identify the zero-mode 
contribution to $L_0$ in (\ref{da5}) with the asymptotic charge (\ref{ab6}), we see 
that the principal continuous representations correspond to black holes, while, as 
noted in section \ref{WGC}, the discrete representations correspond to point particles.
This picture receives further support from the analysis of the Chern-Simons holonomies
of these solutions, which can be related to orbits in $\mathrm{SL}(2,\mathbb{R})$ and
thence to particular representations \cite{Troost}.

We must next consider representations of the full affine algebra (\ref{da1}) of currents
$J^a_n$.  As in standard conformal field theory \cite{CFT}, one can construct a 
representation by starting with a ``vacuum'' $|j,m\rangle$ for which
\beq
J^a_n|j,m\rangle = 0 \qquad \hbox{for $n>0$}
\label{db4}
\eeq
and building a tower of states by acting with current operators $J^a_{-n}$:
\beq
|\psi\rangle = J^{a_1}_{-n_1}J^{a_2}_{-n_2}\dots J^{a_\ell}_{-n_\ell}|j,m\rangle ,
\qquad n_1,n_2,\dots,n_\ell>0 .
\label{db5}
\eeq
The resulting representation spaces are denoted with hats; for example, the
space of states of the form (\ref{db5}) built over a a discrete representation
$\mathcal{D}^+_j$ is $\mathcal{\hat D}^+_j$.  One immediately confronts a serious 
problem: even if the ``vacuum'' states all have positive norm, it is generally easy 
to create negative norm excited states \cite{Balogb,Dixon}.  This in itself is not 
fatal, since one can still require a physical state condition of the form
\beq
L_0|\mathit{phys}\rangle = \alpha |\mathit{phys}\rangle .
\label{db6}
\eeq
(Classically, diffeomorphism invariance requires $\alpha=0$, but as we know
from string theory, there may be quantum corrections.)  A number of authors have
investigated the question of whether one can prove a ``no-ghost theorem'' for
physical states \cite{MalOoga,Hwang,Hwang2,Evans,Petropoulos,Mohammedi,Petropoulosb,%
Asano,Pakman}; the conclusion is that as long as $k>2$ and one restricts to states 
$-k/2\le j<0$ in the discrete representations, one \emph{can} indeed ensure that 
the Hilbert spaces built over $\mathcal{D}^\pm_j$ and $\mathcal{C}^\epsilon_j$ 
are unitary.  Alternatively, it may be possible to use a different free-field 
representation of the WZW model \cite{Bars,Satoh,Hemmb}, in which zero-modes are 
treated quite differently, to achieve a ghost-free spectrum.

We should also demand that when placed on a torus, our WZW model is invariant under 
``large'' diffeomorphisms, diffeomorphisms that cannot be continuously deformed to the
identity.  This is the demand for a modular invariant partition function \cite{CFT}.  
For the representations (\ref{db5}), it seems likely that no such partition function 
is possible \cite{Petropoulosb,Kato}.  But as Henningson et al.\ first pointed out 
\cite{Henn,Hwang2,MalOoga}, the  $\mathrm{SL}(2,\mathbb{R})$ WZW model has additional 
``winding sectors'': for any integer $w$, the transformation
\begin{align}
&J^0_n \rightarrow {\tilde J}^0_n = J^0_n - \frac{k}{2}w\delta_{n,0} , \qquad
J^\pm_n \rightarrow {\tilde J}^\pm_n = J^\pm_{n\pm w}   \nonumber\\
& L_n \rightarrow {\tilde L}_n = L_n + wJ^0_n - \frac{k}{4}w^2\delta_{n,0} , 
\label{db7} 
\end{align}
where $J^\pm = J^1\pm iJ^2$, preserves the commutation relations (\ref{da1})
and generates a new representation labeled by $w$.  Such a transformation is called 
``spectral flow'' in conformal field theory.  For the $\mathrm{SL}(2,\mathbb{R})$ WZW 
model, it has its origin in the fact that $\mathrm{SL}(2,\mathbb{R})$ has a compact 
direction; $w$ corresponds to a winding  number around this $S^1$.  The ``flowed'' 
representations have eigenvalues of $L_0$ that are unbounded below, but one can 
again prove a ``no ghost'' theorem for states obeying (\ref{db6}) \cite{MalOoga}.

The new spectrally flowed sectors (\ref{db7}) are precisely what is needed to achieve 
modular invariance.  Henningson et al.\ show in \cite{Henn} that if one includes all 
integral values of $w$, it is relatively straightforward to write down a modular 
invariant partition function.  (See also \cite{Satoh,Israel}.)  Moreover, the 
result is essentially an analytic continuation of the partition function for the 
$\mathrm{SU}(2)$ WZW model, providing some further justification for the Euclidean 
methods of Ref.\ \cite{Carlipk}.

We now have a candidate for our Hilbert space: the direct sum/integral 
\cite{MalOoga,MalOogc}
\beq
\bigoplus_{w=-\infty}^{\infty}\bigoplus_{j=-k/2}^{-1/2}
  \mathcal{\hat D}^{+,w}_j\bigoplus_{w=-\infty}^{\infty}\bigoplus_{\epsilon,s}
  \mathcal{\hat C}^{\epsilon,w}_{-\frac{1}{2}+is} .
\label{db8}
\eeq
The resulting partition function has been computed in \cite{Israel}.  The question 
now is whether one can count states in this Hilbert space to obtain the BTZ black 
hole entropy.  The answer is not yet known.  Troost and Tsuchiya have pointed out 
two serious problems \cite{Troost}: the spectrum of $L_0$ is not bounded below 
for the spectrally flowed states, and the Hilbert space generally includes negative-norm
states.  Both of these problems can be solved by imposing the physical state
condition (\ref{db6}), but this is presumably \emph{not} what we want to do; we
are interested in counting states with $L_0^\pm$ given by (\ref{ab6}).

Part of the problem may be that we are looking at a pure WZW model, rather than
a WZW model coupled to a background field as in (\ref{cb8}).  If we include  
${\bar A}_z$ in the current as in (\ref{cb8a}), using the BTZ connection 
(\ref{ab7}) as our background field, the zero-mode contribution to $L_0$ changes.  
In fact, we find a shift precisely of the form (\ref{db7}), with a ``winding 
number'' (see also \cite{Hemmb})
\beq
{\bar w} = \frac{r_+\pm r_-}{\ell} .
\label{db9}
\eeq
In general, this $\bar w$ is not an integer---indeed, if it is, then the holonomies
(\ref{ab7a}) are trivial---and the BTZ black hole does not pick out an ordinary 
spectrally flowed sector.  But the background dependence of the WZW action suggests that 
we should perhaps not allow arbitrary spectral flow, but should use the asymptotic
data to restrict the representations we are considering.  This, of course, means
giving up modular invariance for any \emph{particular} black hole, but this may 
not be unreasonable.  In particular, it is known that modular transformations
of the classical Euclidean BTZ black hole do not preserve its physical meaning;
rather, one obtains a family of inequivalent solutions, including hot empty
anti-de Sitter space \cite{MalStrom}.

If we restrict ourselves to the ``winding sector'' determined by $\bar w$, the
physical state condition (\ref{db6}) becomes
\beq
L_0|\mathit{phys}\rangle 
  = \left( - \frac{j(j+1)}{k-2} + N \right)|\mathit{phys}\rangle
  = \left(\alpha - {\bar w}m + \frac{k}{4}{\bar w}^2\right)|\mathit{phys}\rangle .
\label{db10}
\eeq
This gives a value of $L_0$ that agrees, for large $\bar w$, with the classical 
value (\ref{ab6}).  But it leaves us with the dilemma discussed in section \ref{WGC}:  
there do not seem to be enough degrees of freedom in the number operator $N$ to 
account for the Bekenstein-Hawking entropy.  

On the other hand, (\ref{db10}) is  
very similar to the diffeomorphism-invariance condition of \cite{Carlipl}, so it 
could be that a correct combination of left- and right-moving states might yield
the correct entropy.  The similarity with \cite{Carlipl} also suggests that the
conventional boundary conditions might be more easily expressed a different, 
``hyperbolic'' representation of $\mathrm{SL}(2,\mathbb{R})$ \cite{Lindblad}, in 
which the noncompact generator $J_0^2$ is diagonalized.  This is, indeed, the 
natural representation if we wish to view the black hole as a quotient space
of $\mathit{AdS}_3$, since the identification (\ref{ab7a}) is by hyperbolic 
elements of $\mathrm{SL}(2,\mathbb{R})$.  There has been some work on the 
$\mathrm{SL}(2,\mathbb{R})$ WZW model in this context \cite{Satoh,Hemmb,Hemmc,Troostb}, 
but, to the best of my knowledge, no one has yet attempted to count states.

\subsection{Liouville Theory \label{LT}}

The normal boundary dynamics induced from a Chern-Simons theory is described
by a WZW action.  But as we saw in section \ref{WZWL}, slightly stronger 
anti-de Sitter boundary conditions reduce the $\mathrm{SL}(2,\mathbb{R})\times%
\mathrm{SL}(2,\mathbb{R})$ WZW model of (2+1)-dimensional gravity to Liouville
theory.  The central charge (\ref{cc5}) of the Liouville theory has the correct
value to give the BTZ black hole entropy via the Cardy formula.  We might 
therefore hope that a direct counting of states in Liouville theory could
explain this entropy.

The argument is not, unfortunately, quite so straightforward, since the central 
charge in the Cardy formula is the ``effective'' central charge (\ref{bb1}), 
which is not obviously the same as (\ref{cc5}).  Quantum states in Liouville 
theory split into two sectors, ``normalizable'' states and ``nonnormalizable''
or ``Hartle-Hawking'' states.  The names come from the Schr{\"o}dinger picture
description of wave functions: if one writes a wave function $\Psi[\varphi]$
and considers its dependence on the zero-mode $\varphi_0$, the ``normalizable''
states are (delta function) normalizable with respect to the measure $d\varphi_0$ 
\cite{Seiberg}.  The $\mathrm{SL}(2)$-invariant vacuum is not a normalizable 
state, and as a result, the usual operator-state correspondence breaks down:
states created by functionally integrating over a disk with a local operator
insertion are non-normalizable.

Quantum Liouville theory has been studied extensively---for a review, see
\cite{Teschner}---but the emphasis has been largely on the normalizable sector.  
In this sector, the eigenvalues of $L_0$ are bounded below \cite{Seiberg} by
\beq
\Delta_0 = \frac{c-1}{24} .
\label{dc1}
\eeq
As Kutasov and Seiberg have stressed \cite{Kutasov}, this corresponds to an 
effective central charge $c_{\hbox{\scriptsize\it eff}} = 1$, that of a single 
free boson.  This clearly does not give enough states to account for the entropy 
of a BTZ black hole, and has led to the suggestion that Liouville theory is only 
an effective field theory that does not describe the actual degrees of freedom
\cite{Martinecb}.

The nonnormalizable states, on the other hand, have values of $\Delta_0$ 
that go down to zero.  They are, in fact, closely analogous to the point
particle states described in section \ref{WGC} that fill in the gap between
the massless black hole and anti-de Sitter space \cite{BirmCarChen,Chen}. 
If we can formulate a sensible quantum theory that includes these states,
the Cardy formula tells us that we ought to be able to obtain the correct
black hole entropy.
 
The most interesting effort I know of to formulate such a theory is due to
Chen \cite{Chen}.  From the work of Gervais, we know that the Liouville field 
can be written in terms of chiral fields with conformal weights\footnote{Note 
that these are essentially the same as the $\mathrm{SL}(2,\mathbb{R})$ 
zero-modes (\ref{da6}).} \cite{Gervaisb,Gervaise,Gervaisc,Gervaisd}
\beq
\Delta_j = -j - \frac{\tilde{\gamma}^2}{2} j(j+1) , \qquad
\tilde{\gamma}^2 = \frac{c-13 - \sqrt{(c-1)(c-25)}}{6} \approx \frac{12}{c} ,
\label{dc2}
\eeq
whose exchange algebra is that of the quantum group $\mathrm{U}_q(\mathit{sl}_2)$
with $q = e^{\pi i {\tilde\gamma}^2/2}\approx e^{\pi i/k}$.  When $q$ is a root
of unity---classically, when $2k\in\mathbb{Z}$---such a group has only a finite
number of irreducible representations.  A second set of chiral vertex operators 
can be formed by the replacement ${\tilde\gamma}\rightarrow2/{\tilde\gamma}$; 
together, these give a quantum group structure of the form 
$\mathrm{U}_q(\mathit{sl}_2)\odot\mathrm{U}_{\tilde q}(\mathit{sl}_2)$ 
\cite{Gervaisd}.

Following Gervais \cite{Gervaisb,Gervaise,Gervaisc,Gervaisd}, Chen observes that 
the conformal weights (\ref{dc2}) of the Hartle-Hawking states formed by insertions 
of these operators are of the Kac form \cite{CFT}; that is, the states
\beq
|\psi\rangle = L_{-n_1}L_{-n_2}\dots L_{-n_\ell}|\Delta_{j,\tilde j}\rangle
\label{dc3}
\eeq
become singular at level $(2j+1)(2{\tilde j}+1)$. This behavior
occurs in the ``minimal models'' of conformal field theory, where its implications
are well understood \cite{CFT}: 
\begin{enumerate}
\item the singular states are orthogonal to every other state in the representation;
\item the singular states are null, that is, they have zero norm.
\end{enumerate}
Chen argues that the first property continues to hold for Liouville theory, but 
that the second property does not---because of the absence of an $\mathrm{SL}(2)$-%
invariant vacuum, the Ward identities of Liouville theory differ from those of 
standard conformal field theories, and the usual proofs break down. 

If the singular ``decoupling states'' do, in fact, have finite norms, then 
each pair of allowed representations of  $\mathrm{U}_q(\mathit{sl}_2)$ and 
$\mathrm{U}_{\tilde q}(\mathit{sl}_2)$ gives a different, decoupled ``vacuum'' 
upon which one can build ordinary states by acting with operators $L_{-n}$.  If 
$2k$ is an odd integer, the counting of such ``vacua'' is straightforward; one 
finds a total of $12/\tilde{\gamma}^2\approx 6k$ distinct sectors.  Since these 
sectors decouple from each other, the system acts, at least for large $k$, 
like a system of $6k$ scalar fields, with $k$ given by (\ref{aa3a}).  The 
Hardy-Ramanujan expression (\ref{bb4}) for the partition function \cite{Ramanujan},
discussed in section \ref{CF}, then allows us to directly count states.  The
result correctly reproduces the Bekenstein-Hawking entropy (\ref{ab9}) of the 
BTZ black hole.   

Some added support for this picture comes from the close connection between the 
decoupling states' conformal weights, the conformal weights of point particles,
and the quasinormal modes of the BTZ black hole \cite{BirmCarChen}.  The picture
is not at all complete, however.  In particular, the proof that the ``decoupling 
states'' have finite norm is still sketchy, we do not understand the relationship
between these states and the normalizable sector, and we do not know what happens 
when the coupling constant $2k$ is not an odd integer.  Still, I believe that at 
this writing, this is probably the best candidate we have for a genuine conformal 
field theoretic counting of BTZ black hole states.
 
\subsection{Stringy Results \label{SR}}

While the work I have discussed so far can be understood in the context of pure
(2+1)-dimensional gravity, many of the results were inspired by string theory.  As
noted in section \ref{btz}, (2+1)-dimensional anti-de Sitter space is naturally
isometric to $\mathrm{SL}(2,\mathbb{R})$, so string theory on $\mathit{AdS}_3$ 
can be formulated in terms of strings moving on a group manifold.  This topic
was first investigated by Balog et al.\ \cite{Balog}, who discovered the problems
of nonunitarity and ghosts discussed in section \ref{QWZW}.  The idea that the 
BTZ black hole could be treated as an exact string theory background was first 
introduced by Horowitz and Welch \cite{HorWelch} and Kaloper \cite{Kaloper} in 1993, 
and this area has become a minor industry in itself.  

A full description of stringy approaches is too far afield for this review, but 
a few highlights deserve mention.  Section 5 of Ref.\ \cite{AGMOO} has a much
more comprehensive discussion and a good list of references, and I will undoubtedly
omit some important results.\\

\noindent{\bf AdS/CFT Correspondence}: The connection between three-dimensional
asymptotically anti-de Sitter gravity and two-dimensional conformal field theory
is probably the simplest example of Maldacena's celebrated conjecture of a duality
between string theory in asymptotically AdS spacetimes and conformal field theory 
one dimension lower \cite{Maldacenab}.  Strictly speaking, the proposed correspondence 
involves string theory in ten-dimensional asymptotically AdS space, and one should
really consider a product space such as $\mathit{AdS}_3\times S^3\times T^4$.  Many 
of the results we have obtained in the context of pure conformal field theory have 
a simple AdS/CFT interpretation.  For instance, the appearance of the Liouville 
stress-energy tensor in the metric, as described in section \ref{FG}, is exactly 
what one would expect from the AdS/CFT relation between bulk gravitons and boundary 
stress-energy tensors \cite{Gubser,Wittenx,Hyunb}.  The nonlocality of the action
(\ref{cd8}) is also natural: the action is the generator of boundary stress-energy
tensor correlators, and thus cannot be local.  The (2+1)-dimensional model has 
also provided one of the relatively few known time-dependent tests of the AdS/CFT
correspondence, by showing that the response of the boundary conformal field
theory to small perturbations is directly related to the behavior of quasinormal
excitations of the BTZ black hole in the bulk \cite{BirmSachsSolo,Son}.

The AdS/CFT picture also allows us to extend some of these results beyond three
spacetime dimensions.  In \cite{Stromb}, it was shown that for certain higher-dimensional 
near-extremal black holes with near-horizon geometries that looks like that of the 
BTZ black hole, the three-dimensional Cardy formula can again be used to count states.  
This argument has been extended to a wide variety of higher-dimensional stringy black 
holes in, for example, \cite{Hyuna,Sken,Larsen4d,Larsen5d,Larsen4drot,Satohb,Skeny}.  
The connection between quasinormal modes and Liouville states of \cite{BirmCarChen} 
has been extended to these higher-dimensional black holes as well \cite{BirmCar}, 
where such ``stringy'' properties as D-brane charge fractionization have been 
reproduced.\\

\noindent{\bf Behind the Horizon}:  If the AdS/CFT correspondence is correct, the
conformal field theory at the boundary of $\mathit{AdS}_3$ should contain complete 
information about the interior, including information about the interior of any 
event horizons.  At first sight, this seems implausible, since the region inside
a horizon is causally disconnected from the conformal boundary at infinity.  
A number of recent papers have shown, however, that various correlation functions
at infinity can, in fact, probe the interior of a BTZ horizon \cite{BalRoss,%
LoukoRoss,KrausOog,BalLevi,BalLevib}.  Such quantities can be determined in
the Lorentzian theory by analytic continuation from Euclidean values, and with
a suitable choice of contour, a continuation can probe the region behind the
horizon, giving information about the inner horizon and the singularity.  One
can understand these results as coming from the smoothness of the BTZ metric
and the analyticity of field theoretic probes; this analyticity allows us to
determine properties of the metric even in region we cannot directly measure.
It is noteworthy that while these results cannot be obtained in pure gravity%
---one needs a probe such as a scalar field---neither do they depend on the
full apparatus of string theory.\\

\noindent{\bf Stringy Descriptions of the BTZ Black Hole}: Much of the work on
$\mathrm{SL}(2,\mathbb{R})$ WZW models described in section \ref{QWZW} was
motivated by the attempt to understand string propagation on $\mathit{AdS}_3$.
Although they can exist without string theory, the WZW models have features whose
simplest interpretations come from string theory.  For example, the spectrally
flowed sectors (\ref{db7}) in the continuous representation describe ``long strings'' 
that wind around the center of $\mathit{AdS}_3$ \cite{MalOoga}, while the 
restriction $-k/2\le j<0$ required for the no-ghost theorem reflects a ``stringy
exclusion principle'' \cite{MalStrom,Evans}.

A central feature of the string theoretical description is the existence of two
different conformal symmetries, one of the string world sheet and one of the
target space.  In such a model, the Brown-Henneaux central charge (\ref{ba6})
is the central charge of the latter symmetry.  The generators of the spacetime
Virasoro algebra can be constructed in perturbative string theory \cite{GivKutSei,%
deBoer,KutSei}, in a manner that parallels the ``would-be diffeomorphism''
description of section \ref{GCFT}; in particular, the generators of the algebra
correspond to ``pure diffeomorphism'' vertex operators that receive contributions
only from the asymptotic boundary.  Further, at least for the string theory 
defined on $\mathit{AdS}_3\times S^3\times T^4$, it can be shown that $\Delta_0=0$ 
in (\ref{bb1})---at least when $\ell$ is less than the string scale \cite{GKRS}---so 
we need not worry about the ``effective central charge'' dilemma
of section \ref{ECC}.  It should be noted, though, that most of the work in this
area has been on ``Euclidean $\mathit{AdS}_3$'' rather than the real Lorentzian 
sector of the theory.  First steps toward an extension from pure $\mathit{AdS}_3$
to a BTZ black hole background have been taken in \cite{Hemmb,Troost}.

Using D-brane technology, one can also obtain a rather detailed string theoretic
model of gravity on $\mathit{AdS}_3\times S^3\times M^4$, where $M^4$ is 
either a four-torus or a $K3$ manifold.  The description involves a collection 
of D1 branes along a noncompact direction and a collection of D5 branes wrapping 
$M^4$ and sharing the noncompact direction with the D1 branes.  These results are 
reviewed in section 5 of \cite{AGMOO}.  The dual conformal field theory has 
calculable central charge, which again matches the value (\ref{ba6}) obtained 
from pure (2+1)-dimensional general relativity; moreover, a fairly explicit 
description of the degrees of freedom responsible for the black hole entropy is 
now possible.\\

\noindent{\bf CFT and Information Loss}: Pure gravity in 2+1 dimensions cannot 
directly address the ``information loss paradox,'' the question of whether
the formation and subsequent evaporation of a black hole is unitary.  There are,
after all, no propagating degrees of freedom in the purely gravitational theory,
and thus no way for a black hole to evaporate.  If (2+1)-dimensional gravity is
part of a larger string theory, though, the AdS/CFT correspondence implies that
the process should be unitary, since it can be described entirely in terms of an
ordinary conformal field theory.  

This leads to an apparent paradox \cite{Malx}.  The relevant boundary field theory 
lives in a finite volume, and should therefore be subject to Poincar{\'e} recurrences.  
In particular, the correlation functions of small disturbances should be quasiperiodic 
over a long enough time scale.  In the bulk, though, a perturbed black hole relaxes 
exponentially, decaying through quasinormal modes.  The (2+1)-dimensional model, 
in which both quasinormal modes and boundary correlators are exactly computable 
\cite{BirmSachsSolo}, offers a simple arena for investigating this issue.  Proposals 
to resolve the contradiction by summing over bulk topologies apparently fail 
\cite{KlebPorr,Solob}; it has been speculated that one may ultimately have to alter 
the physics near the horizon to eliminate quasinormal excitations \cite{Solob,Barbon}.

\section{What States Are We Counting? \label{what}}
\setcounter{footnote}{0}

Despite some differences among approaches to the boundary dynamics, it seems certain
that asymptotically anti-de Sitter gravity in 2+1 dimensions acquires new degrees of
freedom at the conformal boundary, and that these are associated with a change in
the physical meaning of gauge symmetries at the boundary.  Nemanja Kaloper and
John Terning have suggested a useful analogy \cite{Terning}: if one thinks of the 
boundary as breaking gauge invariance, then the WZW or Liouville fields are 
essentially the Goldstone bosons of this symmetry-breaking, confined to the 
boundary because it is only there that the gauge transformations are not exact 
invariances.  We have also seen that it is plausible, though certainly not proven, 
that we can count the states of these boundary degrees of freedom to obtain the 
Bekenstein-Hawking entropy of the BTZ black hole.  Despite this progress, though, 
deep questions remain about the physical meaning of these degrees of freedom.

One basic problem may be posed as follows.  Suppose one replaces the BTZ black
hole by a (2+1)-dimensional ``star,'' a finite axially symmetric distribution of 
matter.  The exterior metric of such a configuration is still the BTZ metric
\cite{Cruz}, and thus the asymptotic degrees of freedom are identical to those
we have discussed above.  Our state-counting arguments would then apparently
attribute the same entropy to a star as to a black hole.

A perfect fluid is only a phenomenological model of a star, of course, and it 
is possible that a realistic quantum field theory of the constituents would 
alter boundary data and explain the differences in entropy.  This, of course, 
is a basic premise of the AdS/CFT correspondence in string theory \cite{AGMOO}.  
Certainly, new matter fields can lead to modifications to the Fefferman-Graham 
results of section \ref{FG} \cite{Haro,Geg}.  Moreover, as described in section 
\ref{SR}, there is good evidence that such objects as scalar field correlators 
at the conformal boundary can probe the interior of an asymptotically BTZ 
spacetime \cite{BalRoss,LoukoRoss,KrausOog,BalLevi,BalLevib}, perhaps telling 
us whether or not a black hole is present (though see \cite{Marolf}).

Nevertheless, it remains unclear whether we can explain the apparent surplus 
of purely gravitational states in an asymptotically BTZ spacetime containing
no black holes.  Exact (2+1)-dimensional solutions can be found for boson stars 
\cite{Astef}, for example, that have the same asymptotics as the BTZ black hole
but, presumably, different entropies.  Perhaps worse, one can couple scalar fields 
to (2+1)-dimensional gravity in such a way that new exact black hole solutions 
appear \cite{HennMart}, having the same asymptotic symmetries as the BTZ black 
hole, but for which the naive application of the Cardy formula gives the wrong
entropy.  These solutions are probably unstable against decay into ordinary BTZ 
black holes \cite{Geg}, and Park has argued that the correct ``effective central
charge'' (\ref{bb1}) removes the discrepancy \cite{Park}, but it is clear that 
we must proceed with caution.

This problem is closely related to the question of where the degrees of freedom
of the black hole live.  As long as we are considering a single BTZ black hole
in otherwise empty space, the answer to this question is irrelevant, since there
are no ``bulk'' degrees of freedom.  But if we wish to isolate the states of a
black hole in a spacetime in which other matter or fields are present, it would 
be helpful to be able to move the boundary closer to the event horizon.  Similarly,
there are multi-black-hole solutions whose asymptotic symmetries are indistinguishable
from those of a single BTZ black hole \cite{Mansson}; presumably we would like to 
be able to attribute separate entropies to the separate horizons.

In a string theory setting, arguments for including conformal field theories at both 
the conformal boundary and the event horizon have appeared in \cite{Behrndt,Behrndtb}.
For pure (2+1)-dimensional gravity, the question has been investigated in \cite{Carlipg},
taking advantage of the Chern-Simons derivation of the central charge.  Ref.\ 
\cite{Carlipg} starts with the general expression (\ref{ba7a}) for the boundary 
term in the generator of diffeomorphisms and asks the effect of imposing different 
boundary conditions.  For example, we might require that the induced boundary 
metric $g_{\phi\phi}$ be fixed---that is, that the Lie derivative 
$\mathcal{L}_\xi g_{\phi\phi}=0$---at a boundary $r=r_0$.  An easy computation 
\cite{Carlipg,BanOrt} shows that this requires 
\beq
\xi^\rho = - \frac{N(\infty)}{N(r_0)}\partial_\phi\xi^\phi ,
\label{ea1}
\eeq
where $N(r)$ is the BTZ lapse function (\ref{ab2}).  The resulting central
charge can be obtained from (\ref{bc5}): it is
\beq
{\tilde c}(r_0) =
c + \left(\frac{N(\infty)}{N(r_0)}\right)^2 \frac{3\ell}{2G} ,
\label{ea2}
\eeq
significantly different from the conformal boundary case.  Fixing the extrinsic
curvature rather than the metric at the boundary gives a different shift in $c$;
yet another value comes from fixing the mean curvature of the boundary itself
\cite{Carlipg}.  It would thus appear that the location of the boundary is crucial.

In fact, this is not the case.  As stressed in section \ref{ECC}, the relevant
quantity is not the central charge, but the ``effective'' central charge.  In  
each of the cases I have described, the change in the central charge of the
Virasoro algebra is \emph{precisely} canceled by a shift in the minimum conformal
weight $\Delta_0$, leaving the effective central charge $c_{\hbox{\scriptsize\it eff}}$
unaltered.  Thus if we believe the counting arguments of section \ref{CF}, we
have a great deal of latitude in the placement of the boundary upon which the
``would-be gauge'' degrees of freedom live.

For a spacetime containing more than a single black hole, of course, an obvious
choice of boundary is the event horizon itself.  There has been a fair amount of
recent work on the question of whether the entropy of a black hole in any dimension
can be obtained from conformal symmetry or ``would-be gauge'' degrees of freedom at
the horizon---see, for example, \cite{Parkc,Carlipm,Solodukhin,Carlipn,Cadoni,Navarrob,%
Birmb,Gupta,Frolovc,Cvitan,Camblong,Carlipo,Carlipp}---but I believe it is fair to 
say that the results, while intriguing, are still far from conclusive.

\section{Open Questions}
\setcounter{footnote}{0}

It is customary to end a review of this sort with a list of remaining questions.  In
this field, most of the deep questions are still open.  We know that (2+1)-dimensional
gravity induces a two-dimensional conformal field theory at a boundary, including
the conformal boundary of asymptotically anti-de Sitter space.  We know that a good 
way to understand the new degrees of freedom is as ``would-be gauge degrees of 
freedom'' that become dynamical because the boundary changes the meaning of the 
symmetries.  And we know what the central charge of the conformal field theory is, 
and that it (somehow) gives the right state-counting for the BTZ black hole.  Beyond 
that, we have a number of suggestive hints, but we lack any very solid answers.

In particular:
\begin{enumerate}
\item Why does the Cardy formula work so well in giving the entropy of the BTZ
black hole?  How is it that a computation that relies only on classical features%
---classical ``charges'' and the Poisson brackets of classical asymptotic symmetry
generators---gives a correct enumeration of quantum states?  These classical features 
must imprint themselves on the quantum theory in a very fundamental way, determining
basic properties of the Hilbert space, but we do not know how this happens.  Indeed,
we do not yet have a physically intuitive explanation of the Cardy formula itself.
\item Can we construct an appropriate quantum $\mathrm{SL}(2,\mathbb{R})\times%
\mathrm{SL}(2,\mathbb{R})$ WZW model and understand it well enough to count states?
The Cardy formula, naively applied, hints at trouble.  To reproduce the BTZ black 
hole entropy, we will have to do something new, whether introducing (and understanding) 
new vacuum or spectrally flowed sectors, building fractional conformal weight states, 
or, most likely, doing something no one has yet thought of.
\item Can we count states in the nonnormalizable sector of Liouville theory?  Here,
the Cardy formula looks promising, but despite some progress, we remain rather far 
from a complete understanding of the relevant quantum theory.
\item Can we describe states at the horizon, or only at infinity?  If we can only
construct the theory at infinity, how do we distinguish different configurations
with the same asymptotic behavior?
\item Can we couple the ``would-be gauge degrees of freedom'' to matter?  While a
count of these degrees of freedom may give us the BTZ black hole entropy, Hawking
radiation requires something to be radiated, and a quantum theory will be consistent
only if that radiation can react back on the gravitational degrees of freedom.  
Two rather different papers, one looking at the conformal boundary \cite{Emparan} and 
one at the horizon \cite{Hotta}, have begun to address this question, and the results 
of \cite{Emparan} suggest an interesting connection to Hawking radiation.  But so far 
we have only very preliminary results.
\item Do the ``would-be gauge degrees of freedom'' provide the fundamental description
of the states of the quantum BTZ black hole, or are they only effective fields that
reflect a more fundamental underlying theory?  This question may not have a unique
answer---there may well be different quantum theories of gravity, which need not
agree about the source of black hole statistical mechanics---but we do not even
know whether such choices exist.
\item Is the progress we have achieved unique to 2+1 dimensions, or can any of our
results be extended to higher-dimensional spacetimes?
\end{enumerate}

We have a lot of work to do.
Perhaps a future review article will be able to answer some of these questions.

\vspace{1.5ex}
\begin{flushleft}
\large\bf Acknowledgments
\end{flushleft}

I would like to thank Max Ba{\~n}ados, Danny Birmingham, Yujun Chen, 
Stanley Deser, Marc Henneaux, Stephen Hwang, Kirill Krasnov, Kostas
Skenderis, and Sergey Solodukhin for helpful comments.
This work was supported in part by Department of Energy grant
DE-FG02-91ER40674.

\appendix
\section{Conventions \label{App}}

It this appendix, I briefly summarize the conventions used in this paper.  

I use the ``mostly minuses'' or ``west coast'' metric signature, in 
which the Minkowski metric is $\eta_{ab} = \mathit{diag}(1,-1,-1)$.  My 
$\mathrm{SL}(2,\mathbb{R})$ generators are
\beq
T^0 = \frac{1}{2}\left(\begin{array}{cc}0 & 1\\-1 & 0\end{array}\right), \qquad
T^1 = \frac{1}{2}\left(\begin{array}{cc}1 & 0\\ 0 & -1\end{array}\right), \qquad
T^2 = \frac{1}{2}\left(\begin{array}{cc}0 & 1\\1 & 0\end{array}\right) ,
\label{A1}
\eeq
so
\beq
\left[ T^a,T^b\right] = \epsilon^{ab}{}_cT^c 
\label{A2}
\eeq
with $\epsilon_{012} = \epsilon^{012} = 1$, and
\beq
{\hat g}^{ab} = \mathrm{Tr}(T^aT^b) = -\frac{1}{2}\eta^{ab} .
\label{A3}
\eeq
My curvature tensor conventions are
\beq
\left[ \nabla_\mu,\nabla_\nu \right]v^a = R_{\mu\nu}{}^a{}_bv^b 
\label{A4}
\eeq
with
\beq
R_{\mu\rho} = R_{\mu\nu\rho}{}^\nu .
\label{A5}
\eeq

\end{document}